\documentclass[aps,prc,letterpaper,11pt,twoside,tightenlines,nofootinbib,showpacs,preprint]{revtex4}
\usepackage{graphicx}
\usepackage[sort&compress]{natbib}
\begin{document}

%%%   New Definitions
%\newcommand{\eg}{{\it e.g.}}
%\newcommand{\etal}{{\it et. al.}}
%\newcommand{\ie}{{\it i.e.}}
\newcommand{\be}{\begin{equation}}
\newcommand{\ee}{\end{equation}}
\newcommand{\bea}{\begin{eqnarray}}
\newcommand{\eea}{\end{eqnarray}}
\newcommand{\bef}{\begin{figure}}
\newcommand{\eef}{\end{figure}}
\newcommand{\bce}{\begin{center}}
\newcommand{\ece}{\end{center}}

\title{$T-$Matrix Approach to Quarkonium Correlation Functions in the QGP}

\author{D.~Cabrera}
\affiliation{Cyclotron Institute and Physics Department, Texas A\&M
University, College Station, Texas 77843-3366, U.S.A.}
\author{R.~Rapp}
\affiliation{Cyclotron Institute and Physics Department, Texas A\&M
University, College Station, Texas 77843-3366, U.S.A.}

\date{\today}

\begin{abstract}
We study the evolution of heavy quarkonium states with temperature in a
Quark Gluon Plasma (QGP) by evaluating the in-medium $Q$-$\bar{Q}$ $T-$matrix 
within a reduced Bethe-Salpeter equation in both $S-$ and $P-$wave
channels. The underlying interaction kernel is extracted from recent 
finite-temperature QCD lattice calculations of the singlet free energy 
of a $Q$-$\bar{Q}$ pair.   
The bound states are found to gradually move above the $Q$-$\bar{Q}$
threshold after which they rapidly dissolve in the hot system.
The $T-$matrix approach is particularly suited to investigate these 
mechanisms as it provides a unified treatment of bound and 
scattering states including threshold effects and the transition to
the (perturbative) continuum.
We apply the $T-$matrix to calculate $Q$-$\bar{Q}$ spectral 
functions as well as pertinent Euclidean-time correlation functions 
which are compared to results from lattice QCD. 
A detailed analysis reveals large sensitivities to the interplay 
of bound and scattering states, to
temperature dependent threshold energies and to the ``reconstructed'' correlator
used for normalization.  
We furthermore investigate the impact of finite-width effects on the 
single-quark propagators in the QGP as estimated from recent
applications of heavy-quark rescattering to RHIC data.
\end{abstract}

\maketitle

%%%%%%%%%%%%%%%%%%%%%%%%%%%%%%%%%%%%%%%%%%%%%%%%%%
\section{Introduction}
\label{sec:intro}
%%%%%%%%%%%%%%%%%%%%%%%%%%%%%%%%%%%%%%%%%%%%%%%%%%
Bound states of heavy quarks (charm and bottom, $Q$~=~$c$,~$b$) have 
long been recognized as valuable objects for spectroscopy in
Quantum Chromodynamics (QCD), thereby illuminating the nature of the 
static quark-antiquark potential (cf.~Ref.~\cite{Brambilla:2004wf} 
for a recent comprehensive overview).   
This opportunity carries over when embedding quarkonia into hot and/or
dense matter, providing a rich laboratory for the study of medium 
modifications. The latter include (Debye-) color-screening of the 
$Q$-$\bar Q$ interaction, dissociation reactions induced by constituents 
of the medium, and the change in thresholds caused by mass (or width) 
modifications of open heavy-flavor states ($D$ and $B$ mesons or $c$ 
and $b$ quarks). The challenge is to develop a theoretical framework 
that allows a comprehensive description of heavy quarkonia in the 
Quark-Gluon Plasma and their production in ultrarelativistic
heavy-ion collisions.  

Lattice QCD (lQCD) calculations have made substantial progress in 
characterizing in-medium quarkonium properties from first principles. 
In particular, it has been found that ground state 
charmonia~\cite{Asakawa:2003re,Datta:2003ww,Umeda:2002vr} 
and bottomonia~\cite{Petrov:2005ej} do not dissolve until significantly
above the critical temperature, $T_c$.
This finding has been qualitatively supported in model  
calculations based on potentials extracted from lQCD, using either 
a Schr\"odinger equation to solve the bound state 
problem~\cite{Shuryak:2003ty,Wong:2004zr,Alberico:2005xw,Mocsy:2005qw},
or a $T$-matrix approach which simultaneously accounts for scattering 
states~\cite{Mannarelli:2005pz}. The survival of low-lying quarkonia
above $T_c$, in connection with effects of color-screening, 
parton-induced dissociation and medium modified open-charm and -bottom 
thresholds, has recently been implemented for heavy-ion 
collisions~\cite{Grandchamp:2003uw,Grandchamp:2005yw}.

A more quantitative (and reliable) comparison of model calculations 
to lQCD can be performed
at the level of (spacelike) Euclidean-time correlation 
functions~\cite{Rapp:2002pn,Mocsy:2004bv}. 
The latter are directly evaluated in lQCD with good accuracy, while 
the conversion of (timelike) spectral functions as evaluated in model 
approaches merely involves a straightforward convolution with a thermal 
weight function (as opposed to an inverse integral transform when going 
from Euclidean to Minkowski space). 
One of the challenges in such studies is that the model calculations 
need to describe not only the bound-state part of the spectral function
but also its continuum part as well as threshold effects. 
In Ref.~\cite{Mocsy:2005qw} a quantitative calculation of Euclidean
correlators was performed using temperature-dependent heavy-quark 
potentials in a Schr\"odinger equation. The latter has been used to 
determine the bound-state spectrum in $\delta$-function approximation 
(characterized by a binding energy and amplitude (or decay constant)), 
while the (onset of the) continuum was approximated
with perturbation theory. While general trends of 
the lQCD correlators were captured, significant discrepancies 
were established especially in the $S$-wave charmonium channels 
($\eta_c$ and $J/\psi$). In particular, the importance of a reliable
treatment of the continuum threshold was recognized.
 
In the present paper we evaluate charmonium and bottomonium correlators
using a different method. The basic input are still in-medium 
$Q$-$\bar Q$ potentials as estimated from lQCD, but we will employ these 
within a scattering equation to calculate the in-medium $Q$-$\bar Q$ 
$T-$matrix~\cite{Mannarelli:2005pz}. The main advantage of the 
$T-$matrix approach is that it simultaneously incorporates bound and 
scattering states based on the same interaction. 
Especially for situations of dissolving bound states (as expected for 
the problem at hand) the $T-$matrix provides a more 
comprehensive, and thus more reliable, description of the underlying 
nonperturbative effects. At the correlator level, the high-energy limit 
can be recovered by appropriate normalization of the uncorrelated 
(perturbative) limit, and no decomposition into bound-state and 
continuum parts is necessary. In addition, the $T-$matrix equation 
allows for a straightforward implementation of in-medium single-particle 
(quark) properties via pertinent self-energy insertions in the 
two-particle Green's function, which we will also investigate.     

Our article is organized as follows: in Sec.~\ref{sec:BS} we recall 
the basic set-up of, and input to, the two-body scattering equation, 
including partial-wave expanded potentials and single-quark 
selfenergy insertions.
In Sec.~\ref{sec:Tmat} we evaluate the finite-temperature $T-$matrices 
for $S-$ and $P-$wave quarkonia; we first extract heavy-quark potentials 
from lQCD in Sec.~\ref{ssec:Lat}, including a discussion of its 
short- and large-distance limits and relations to single-quark 
properties, followed by our baseline results for the finite-temperature 
quarkonium $T-$matrices in Sec.~\ref{ssec:Tmat1}.   
In Sec.~\ref{sec:Corr} the latter are first employed to construct 
pertinent spectral functions (Sec.~\ref{ssec:Spec}), followed by a 
calculation of Euclidean correlators (Sec.~\ref{ssec:Eucl}) and 
a discussion of their properties in comparison to other model
and lQCD results (Sec.~\ref{ssec:Cres}).
Sec.~\ref{sec:Concl} contains our conclusions and an outlook. 

%%%%%%%%%%%%%%%%%%%%%%%%%%%%%%%%%%%%%%%%%%%%%%%%%%%%%%%%%%%%%%%%%%%
\section{Scattering equation and identification of bound states}
\label{sec:BS}
%%%%%%%%%%%%%%%%%%%%%%%%%%%%%%%%%%%%%%%%%%%%%%%%%%%%%%%%%%%%%%%%%%%
We here summarize the main features of the $T-$matrix approach to 
study quark-antiquark interactions in the QGP, as employed in 
Ref.~\cite{Mannarelli:2005pz}. It utilizes a 
three-dimensional reduction of the Bethe-Salpeter equation which 
neglects virtual particle-antiparticle loops and amounts to resumming 
the scattering series in ladder approximation. The pertinent
Lippmann-Schwinger equation for the off-shell $T-$matrix in a given 
partial-wave channel (specified by angular momentum $l$) 
reads\footnote{The partial wave expansion reads 
$T=4\pi \sum_l (2l+1)\, T_l\,P_l(\cos\theta)$, and similarly for
the potential.}
\be
\label{LS}
T_l(E;q',q) = V_l(q',q) + \frac{2}{\pi} \int_0^{\infty} dk \, k^2 \, 
V_l(q',k)\, G_{\bar{Q}{Q}}(E;k) \, T_l(E;k,q) \,[1-2f^Q(\omega_k)] \ ,
\ee
where $q$ ($q'$) are the incoming (outgoing) relative quark  three-momenta in
the center of mass (CM) frame and $E$ is the CM  energy\footnote{The on-shell
$T-$matrix is defined for $q=q'$ and  $E=\sqrt{s}=2\,\omega_q$, with $\omega_q$
the (relativistic) on-shell heavy-quark energy.}. 
Eq.~(\ref{LS}) is written for vanishing total 3-momentum of the 
heavy-quark pair, which gives the above (simple) form of the Pauli 
blocking factor with $f^Q(\omega)=[\exp(\omega/T)+1]^{-1}$. The 
intermediate two-particle propagator is evaluated in the 
Blankenbecler-Sugar reduction 
scheme~\cite{Blankenbecler:1965gx} 
(uncertainties due
to other reduction schemes have been checked to be  
small~\cite{Mannarelli:2005pz}), 
\be
\label{BbS}
G_{\bar{Q}{Q}}(E;k) = \frac{m^2}{\omega_k} \ 
\frac{1}{s/4 - \omega_k^2 - 2i\,\omega_k
\textrm{Im}\,\Sigma(\omega_k,k)} \ ,
\ee
where $\omega_k$ is the solution of the quark dispersion relation,
\be
\label{disp}
\omega_k = \sqrt{m^2+k^2} + \textrm{Re}\,\Sigma(\omega_k,k) \ , 
\ee
with a quark-mass term ($m$) and selfenergy ($\Sigma$) to be 
discussed below.
The interaction kernel of the scattering equation, $V_l(q',q)$, is 
provided by the heavy-quark potential in momentum space. It follows 
from a Fourier transformation of the 
coordinate-space potential, $V(r)$, which we obtain
from lQCD calculations as elaborated in Sec.~\ref{ssec:Lat} below. 
The components of the potential in the partial-wave basis are given by
\be
\label{pot-Fourier-projected}
V_l(q',q) = \frac{1}{8\pi} \int_{-1}^{+1} du_{q'q} V(\vec{q}\,',\vec{q})\,
P_l(u_{q'q})
= \frac{1}{8\pi} \int_{-1}^{+1} du_{q'q} P_l(u_{q'q}) 
\int d^3r \, V(r) \, e^{i\,(\vec{q}-\vec{q}\,')\vec{r}} \,\,\,,
\ee
with $P_l(x)$ the Legendre polynomial of degree $l$ and
$u_{q'q}=\cos <\widehat{\vec{q},\vec{q}\,'}>$.

The $T-$matrix equation (\ref{LS}) is solved with the algorithm of 
Haftel and Tabakin~\cite{Haftel:1970}: after discretizing the momentum
integration, Eq.~(\ref{LS}) is converted into a matrix equation,
\be
\label{HT1}
\sum_{k=1}^N {\mathcal F}(E)_{ik} \,  T(E)_{kj} =
V_{ij} \,\,\, ,
\ee
where, schematically,  ${\mathcal F}=1- w~V~G_{\bar{Q}{Q}}~[1-2\,f^Q]$ 
(with $w$ 
denoting an integration weight). The solution for the $T-$matrix then
follows from matrix inversion. 

To assess the presence of heavy quark-antiquark bound states, the
$T-$matrix has to be studied below the $Q$-$\bar{Q}$ threshold, 
$E_{th}=2 \, \omega_{q=0}$. The non-relativistic potential, 
$V_l(q',q)$, is only defined for real external three-momenta, and 
therefore an evaluation of $T$ below the $Q$-$\bar{Q}$ threshold 
requires a prescription for the subthreshold continuation of the 
potential.
For $S-$wave scattering we follow the standard convention of
setting the momenta to zero~\cite{Aouissat:1994sx}, 
\begin{equation}
\label{T-Swave}
T_0(E<E_{th}) = T_0(E;q'=q=0)  \ .
\end{equation}
The reliability of this continuation can be checked by exploiting 
the (numerical) matrix form of the scattering equation. Since a bound 
state corresponds to a pole of the amplitude on the real energy axis 
below threshold, it follows that the determinant of the transition 
matrix ${\cal F}$ must vanish at the bound-state 
energy~\cite{Haftel:1970},  
\be
\label{boundstate}
\det {\mathcal F}(E) = 0 \  , \  E < E_{th}  \ .
\ee
A similar condition arises from the solution of the Schr\"odinger 
equation for the bound state problem~\cite{Haftel:1970,ropke}. This is 
equivalent to finding the zeroes of the Jost function in scattering 
($S-$matrix) theory.
For $P-$wave states, the potential is proportional to the in- and
outgoing quark momentum and therefore the continuation in Eq.~(\ref{T-Swave}) 
cannot be applied. However, the condition Eq.~(\ref{boundstate})
remains valid and will be used to determine $P-$wave bound
states. 

The quark selfenergy figuring into the two-particle propagator, 
Eq.~(\ref{BbS}), receives contributions from
interactions with (light) anti-/quarks and gluons in the heat
bath. In Ref.~\cite{Mannarelli:2005pz} this was schematically
written as
\be
\label{qself}
\Sigma = \widetilde{\Sigma} + \int f^q~T_{Qq}~S_q \ , 
\ee
where $\widetilde{\Sigma}$ denotes the gluonic piece and the second
term involves the heavy-light quark $T-$matrix closed by a light-quark
propagator, $S_q$, and a thermal distribution, $f^q$. Rather than using 
an explicit
model calculation for $T_{Qq}$~\cite{Mannarelli:2006}, in the present
work we will constrain ourselves to the following levels of
approximation:
(a) a fixed heavy-quark mass $m$ (i.e., $\textrm{Re}\,\Sigma=0$) together
   with a small imaginary part, $\textrm{Im}\,\Sigma=-$0.01~GeV, mostly
   for numerical purposes (to avoid $\delta$-function like
   bound states in the $T-$matrix);
(b) a temperature dependent heavy-quark mass as estimated from
    the asymptotic value of the lQCD heavy-quark internal energies;
(c) a heavy-quark width as calculated in an effective model for 
    resonance (plus perturbative gluon) interactions in the 
    QGP~\cite{vanHees:2004gq}, which has been shown to give
    reasonable agreement with data on suppression and elliptic flow
    of semileptonic electron spectra from heavy-quark decays
    in Au-Au collisions at RHIC~\cite{vanHees:2005wb}.  
We also note that interactions with heavy anti-/quarks from the
medium can be safely neglected due to the smallness of the number
of $Q$'s in the system. This is different to (and simpler than) the
situation of the light-quark selfenergy which, in turn,
figures into the calculation of the $T-$matrix, 
constituting a self-consistency problem as has been evaluated, e.g.,
in Ref.~\cite{Mannarelli:2005pz}.

%%%%%%%%%%%%%%%%%%%%%%%%%%%%%%%%%%%%%%%%%%%%%%%%%%%%%%%%%%%%%%%%%%%
\section{Temperature Evolution of Heavy Quarkonium $T-$Matrices}
\label{sec:Tmat}
%%%%%%%%%%%%%%%%%%%%%%%%%%%%%%%%%%%%%%%%%%%%%%%%%%%%%%%%%%%%%%%%%%%

%%%%%%%%%%%%%%%%%%%%%%%%%%%%%%%%%%%%%%%%%%%%%%%%%%%%%%%%%%%%%%%%%%
\subsection{Quark-Antiquark Potential from Lattice QCD}
\label{ssec:Lat}
%%%%%%%%%%%%%%%%%%%%%%%%%%%%%%%%%%%%%%%%%%%%%%%%%%%%%%%%%%%%%%%%%%
For the driving kernel of the scattering equation we focus 
on heavy-quark potentials from lQCD.
There is an ongoing discussion as to how to properly extract them from
the (static) heavy-quark free energy. Using directly the latter leads to 
a dissociation temperature of ground-state charmonia of about 
$T_{diss}\simeq1.1~T_c$~\cite{Digal:2001ue}, while the lattice analysis 
of spectral functions suggests that $\eta_c$ and $J/\psi$ survive 
up to around $2\,T_c$. These higher values for $T_{diss}$ can be recovered
if the (color-singlet) internal energy,
\be
\label{U1}
U_1 = F_1 - T \, \frac{dF_1}{dT} \ ,
\ee
is identified with the $Q$-$\bar{Q}$ 
potential~\cite{Wong:2004zr,Mannarelli:2005pz,Alberico:2005xw,Mocsy:2005qw}.
Another possibility, namely a suitable linear combination of $U_1$ and
$F_1$, has been suggested in Ref.~\cite{Wong:2004zr}. The temperature
derivative of  discrete lQCD ``data points'' involved in the extraction of $U_1$
induces  significant uncertainty which is comparable to, e.g., the difference
between quenched and unquenched results (after rescaling of the critical
temperature), as studied in Ref.~\cite{Mannarelli:2005pz}.  In view of this
situation, one has to accept a certain level of  uncertainty in the potential.
To adequately account for this, we adopt  two versions of the internal energy,
Eq.~(\ref{U1}), as the potential: (i) based on fits to the two-flavor 
lQCD results for the free energy from 
Refs.~\cite{Kaczmarek:2003dp,Petreczky:2004priv}, we explicitely perform 
the temperature derivative in Eq.~(\ref{U1}); 
(ii) we directly fit the two-flavor lQCD internal energy data as 
extracted in the calculation of Ref.~\cite{Kaczmarek:2005gi}. Further 
investigations of the impact of using different definitions of the 
$Q$-$\bar{Q}$ potential will be considered in future work.

The long-distance limit of the potential entering Eq.~(\ref{LS}) has to 
be normalized to zero to ensure the convergence of the scattering equation,
\be
\label{UminusUinf}
V(r,T) = U_1(r,T) - U_1^{\infty}(T) \ ,
\ee
with $U_1^{\infty}(T) \equiv U_1(r\to\infty,T)$. 
In Refs.~\cite{Wong:2004zr,Alberico:2005xw,Mocsy:2005qw,Satz:2006uh,Wong:2006bx} 
the linearity of the Schr\"odinger equation is exploited to
trade the internal energy at infinite distance into the energy
of the bound states. In Ref.~\cite{Mocsy:2005qw}, $U_1^{\infty}(T)$ is
interpreted as an effective in-medium contribution to the quark mass, $\Delta
m_Q (T) = U_1^{\infty}(T)/2$ and implemented as a change in the
$Q$-$\bar{Q}$ threshold energy, $E_{th} = 2\,m_Q + U_1^{\infty}(T)$, in the
calculation of the mesonic spectral function.
It is argued that this correction should not modify the mass operator in the
Schr\"odinger equation since quarks inside a bound state do not ``sense" the
medium and therefore should not be subject to medium-induced mass modifications.
This ambiguity in the interpretation of the internal energy at infinite 
distance can be resolved within the many-body scattering equation approach. The
interaction of the quark with the surrounding medium induces a selfenergy which
is encoded in an effective mass change ($\Delta m_Q = \textrm{Re}\,\Sigma$). 
As such, this medium effect has to be included in the two-particle propagator,
Eq.~(\ref{BbS}), and therefore contributes in a nonlinear way as it 
is iterated to all orders in the scattering equation series. As we
discuss in the following sections, this not only modifies the $Q$-$\bar{Q}$
threshold energy but the evolution of the binding energy (total mass of the
bound state) with temperature, leading to different dissociation
temperatures when the effective in-medium mass is considered.

Another intriguing problem regarding the interpretation of $U_1^{\infty}$ as an
in-medium quark mass (or selfenergy) is its possible momentum dependence. Whereas
at infinite separation (low momentum transfer) $U_1^{\infty}$ operates as
$\Delta m_Q$, at short distances (high momentum transfer) the $Q$-$\bar{Q}$
system is
no longer sensitive to the medium, and the quark mass should be unaffected.
Implementing such a momentum dependent quark selfenergy would require a
microscopic treatment of the mass subtraction of the internal energy. 
Instead, we will consider two limiting scenarios, namely 
(a) no in-medium mass correction and (b) in-medium effective
mass as given by $\Delta m_Q (T) = U_1^{\infty}(T)/2$.
  
Finally, for the potentials in momentum space we introduce a 
relativistic correction motivated by the velocity-velocity (Breit) 
interaction in  electrodynamics~\cite{Brown:1952ph}, see also  
Ref.~\cite{Mannarelli:2005pz}. It amounts to the following factor:
\be
\label{Breit}
V(q',q) \to V(q',q) \, 
[1+q'^2/\omega_{q'}^2]^{1/2}\,[1+q^2/\omega_q^2]^{1/2} \ .
\ee

\begin{center}
\begin{figure}[!th]
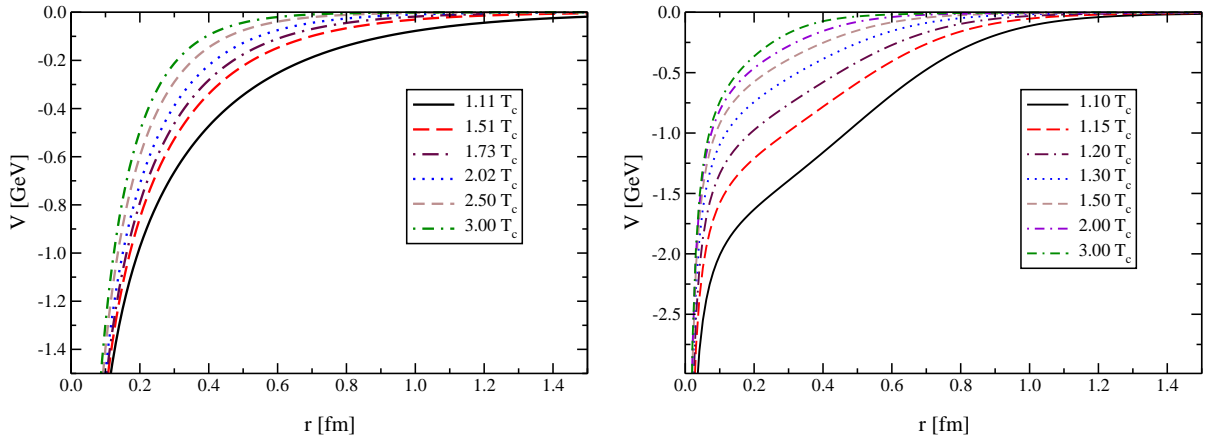

\includegraphics[width=0.47\textwidth]{pot.eps}
\hspace{.2cm}
\includegraphics[width=0.47\textwidth]{potKZ.eps}
\caption{$Q$-$\bar{Q}$ color-singlet potential for several temperatures 
above $T_c$, as defined in Eq.~(\ref{UminusUinf}); left panel: based 
on $F_1$ from Refs.~\cite{Kaczmarek:2003dp,Petreczky:2004priv} in 
connection with Eq.~(\ref{U1}); right panel: based on $U_1$ as directly
evaluated in lQCD in Ref.~\cite{Kaczmarek:2005gi}.}
\label{fig:V1}
\end{figure}
\end{center}

The two potentials used in this work are summarized in Fig.~\ref{fig:V1}.
For case (i) discussed above (extraction from the lQCD free energy),
it evolves rather smoothly with temperature (left panel), while for case 
(ii) (lQCD internal energy) the potential is initially more attractive but 
weakens rapidly with temperature (right panel) for $T\lesssim 1.5\,T_c$ and
slows down thereafter. The rapid decrease is mostly induced by 
$U_1^{\infty}(T)$, see also Ref.~\cite{Kaczmarek:2005gi}.

%%%%%%%%%%%%%%%%%%%%%%%%%%%%%%%%%%%%%%%%%%%%%%%%%%%
\subsection{Quarkonium $T-$Matrices in the QGP}
\label{ssec:Tmat1}
%%%%%%%%%%%%%%%%%%%%%%%%%%%%%%%%%%%%%%%%%%%%%%%%%%%
We now turn to the numerical results for the finite-temperature 
$T-$matrices in the $c$-$\bar{c}$ and $b$-$\bar{b}$ sectors, obtained by 
solving the scattering equation~(\ref{LS}) in both $S-$ and $P-$wave
channels as described above. 

%%%%%%%%%%%%%%%%%%%%%%%%%%%%%%%%
\subsubsection{$S$-Wave States}
\label{sssec_swave}
%%%%%%%%%%%%%%%%%%%%%%%%%%%%%%%
In a first step, we consider the case of narrow quark spectral functions
with $\textrm{Im}\,\Sigma = -10$~MeV (for numerical purposes) and 
constant (temperature-independent) heavy-quark masses 
($\textrm{Re}\,\Sigma = 0$). The latter
are fixed  so that the corresponding ground states are located 
approximately at their vacuum masses for the lowest considered 
temperature ($T=1.1\,T_c$), yielding $m_c = 1.7$~GeV and $m_b = 5.15$~GeV.
\begin{center}
\begin{figure}[!th]
\includegraphics[width=0.8\textwidth]{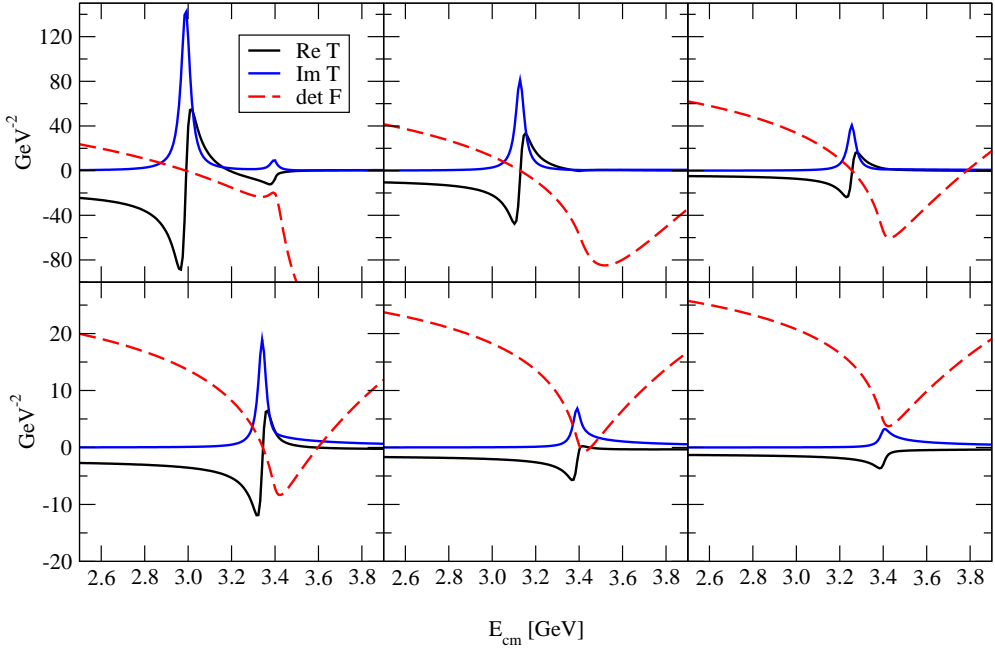}
\caption{Real and imaginary parts of the $T-$matrix for $S-$wave $c$-$\bar{c}$
scattering in the QGP based on potentials derived from the lQCD
free energy of Ref.~\cite{Petreczky:2004priv}.
Also shown is the determinant function $\det {\cal F}$ (dashed line,
arbitrary units). From left to right and up to down the temperatures are
$(1.1,1.5,2.0,2.5,3.0,3.3)\,T_c$.}
\label{fig:TccS}
\end{figure}
\end{center}
Fig.~\ref{fig:TccS} summarizes the on-shell $S-$wave $c$-$\bar{c}$ 
scattering amplitude as a function of CM energy, for several 
temperatures from $1.1 \, T_c$ to $3.3 \, T_c$, as well as the 
determinant function $\det {\mathcal F}(E)$ (in arbitrary units). Since 
we do not include the hyperfine (spin-spin) interactions,  
$\eta_{c}$ ($\eta_b$) and $J/\psi$ ($\Upsilon$) states are 
degenerate. At the lowest temperature, we recover the charmonium ground 
state at $E \approx 3.0$~GeV, and also find a cusp at the $c\bar c$ threshold 
energy indicating that the first excited state (ascribed to the $\psi'$) 
has just melted. 
The determinant function, $\det {\mathcal F}(E)$, vanishes
exactly at the ground state energy coinciding with $\text{Re}\,T(E)=0$, 
thus corroborating our subthreshold continuation of the $T-$matrix,
Eq.~(\ref{T-Swave}). For higher energies $\det {\mathcal F}(E)$ 
approaches zero again, but as soon as the threshold
is reached it deviates indicating that the first excited state has already
crossed into the continuum spectrum\footnote{Strictly speaking, 
${\mathcal F}(E)$ is not a purely real 
function since we have included a small imaginary part in the 
two-particle propagator. 
More precisely, the bound-state condition, Eq. (\ref{boundstate}), 
reads $\textrm{Re}\lbrace \det {\mathcal F}(E) \rbrace = 0$. We are 
plotting $\textrm{Re}\lbrace \det {\mathcal F}(E) \rbrace$ in 
Figs.~\ref{fig:TccS}, \ref{fig:TbbS} and \ref{fig:detF-P}.}.
As the temperature is increased, the bound charmonium state gradually 
moves toward threshold, indicating a reduction of its binding energy.
At the same time, the magnitude of the $T-$matrix is appreciably 
reduced. The $J/\psi(1S)$ survives as a bound state well beyond $T_c$,
eventually crossing the threshold at about $(2.8-3.0)\, T_c$, after which
it turns into a resonance and rapidly
melts in the hot system\footnote{We 
refer to a state ``melting" or ``dissolving" when the scattering 
amplitude is strongly broadened and diminished 
corresponding to a loss of the resonant structure.}.
Our results agree reasonably well with those of Ref.~\cite{Mannarelli:2005pz},
with some differences in size and shape of the scattering amplitude,
in particular a larger dissociation temperature. This is mostly
due to a different parametrization of the $Q$-$\bar{Q}$
potential (reflecting the uncertainties in a derivation of the $Q$-$\bar{Q}$
internal energy from a fit to the free energy, cf.~Sec.~\ref{sssec_pot} 
below) and a different choice of the two-particle propagator, 
which implies deviations at order ${\cal O}(p/m)$. The robustness of 
this approach to dynamically generate quarkonium bound and scattering 
states and their evolution with temperature is confirmed and extended 
in the following to study other quarkonium states.

\begin{center}
\begin{figure}[!tb]
\includegraphics[width=0.8\textwidth]{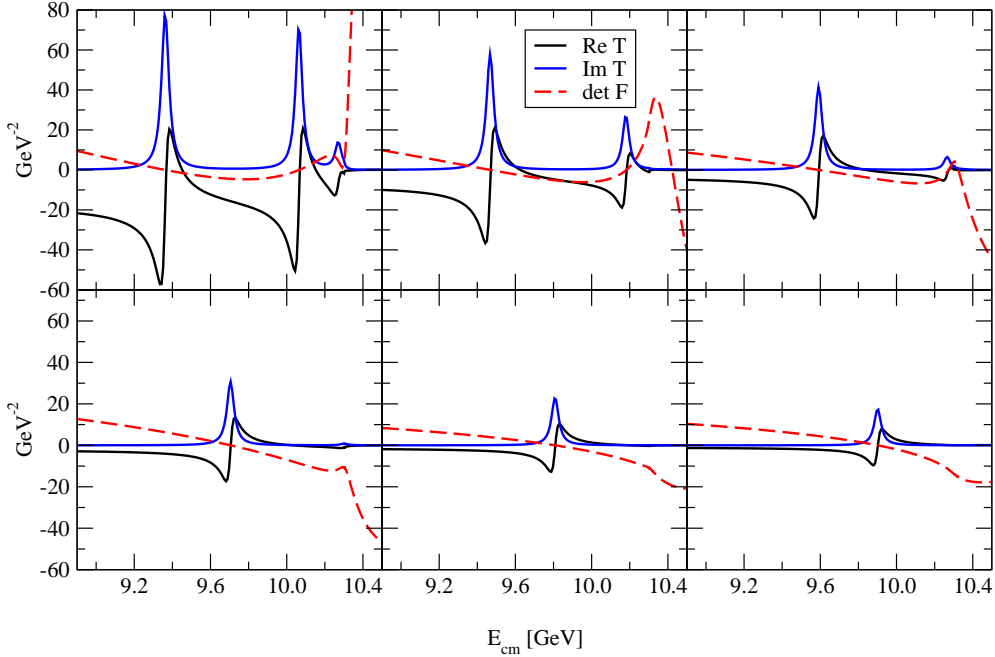}
\caption{Same as in Fig.~\ref{fig:TccS} but for $S-$wave $b$-$\bar{b}$
scattering. From left to right and up to down the temperatures are
$(1.1,1.5,1.8,2.1,2.7,3.5)\,T_c$.}
\label{fig:TbbS}
\end{figure}
\end{center}

The results for $S-$wave $b$-$\bar b$ scattering are depicted in 
Fig.~\ref{fig:TbbS}. At the lowest temperature the $T-$matrix exhibits 
two bound states, as well as the remnant of a third one. The 
bound-state locations are again quantitatively confirmed by the vanishing 
determinant of the transition matrix (dashed lines), while it barely  
reaches zero at the location of the third structure in the $T-$matrix, 
which carries much smaller strength, indicating that it has practically 
melted in the medium. 
The two bound states at $E \approx 9.35,\, 10.05$~GeV are ascribed to  
the ground and first-excited bottomonium states $\Upsilon(1S)$,
$\eta_b$ and $\Upsilon(2S)$, $\eta_b'$, respectively. 
The $\Upsilon(2S)$ 
moves above the $b$-$\bar{b}$ threshold at $T \approx 1.8 \, T_c$, 
whereas the $1S$ state survives in the QGP until much higher 
temperatures, beyond $T \approx 3.5 \, T_c$.

%%%%%%%%%%%%%%%%%%%%%%%%%%%%%%%%
\subsubsection{$P$-Wave States}
\label{sssec_pwave}
%%%%%%%%%%%%%%%%%%%%%%%%%%%%%%%
\begin{center}
\begin{figure}[!tb]
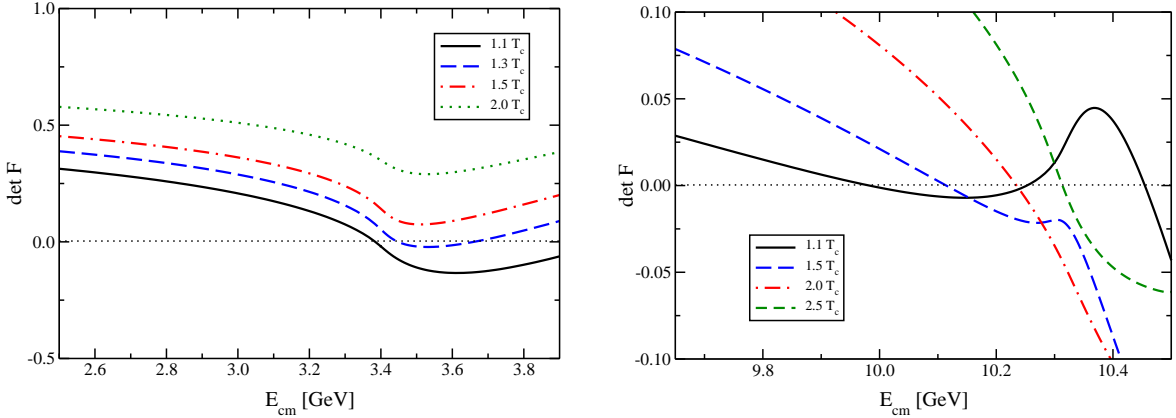

\includegraphics[width=0.45\textwidth]{detFccP.eps}
\hspace{0.5cm}
\includegraphics[width=0.45\textwidth]{detFbbP.eps}
\caption{Pole identification function ($\det {\cal F}$) for charmonium (left)
and bottomonium (right) $P-$wave scattering at several temperatures.}
\label{fig:detF-P}
\end{figure}
\end{center}
Next we study $Q$-$\bar{Q}$ scattering in a relative 
$P-$wave. In order to assess the formation of bound states in this 
channel we rely on the condition in Eq.~(\ref{boundstate}) and 
determine the zeroes of $\det {\cal F}(E)$, which is plotted in 
Fig.~\ref{fig:detF-P} for several temperatures for both charm and bottom.
We only find one $c$-$\bar c$ bound state at the lowest temperature
($T=1.1\,T_c$), at $E \approx 3.4$~GeV (just below threshold), which we 
associate with the $1P$ charmonium, $\chi_c$. As the temperature
increases, the $\chi_c$ state rapidly shifts into the continuum.

The $P-$wave $b$-$\bar{b}$ system exhibits two bound states at the lowest
temperature, which we may identify with the $\chi_b(1P)$ and $\chi_b(2P)$ 
as their energies ($E=9.95,10.25$~GeV) are close to the 
nominal values in the vacuum. The $\chi_b (2P)$ state moves beyond 
threshold for $T\approx 1.3\,T_c$ and the $(1P)$ state for 
$T\approx 2.3\,T_c$. Both the mass and the binding energies, 
$(E_B=E_{th}-M)$, of the $P-$wave states are summarized in 
Tab.~\ref{Ptable} for several temperatures.

\begin{center}
\begin{table}
\begin{tabular}{c||ccccc}
$T/T_c$ & 1.1 & 1.3 & 1.5 & 2 & 2.3\\
\hline
\hline
$M[\chi_c(1P)]$ & 3.38 & - & - & - & - \\
\hline
$E_B[\chi_c(1P)]$ & $\approx$ 0 & - & - & - & - \\
\hline
\hline
$M[\chi_b(1P)]$ & 9.95 & 10.05 & 10.11 & 10.23 & 10.30 \\
\hline
$E_B[\chi_b(1P)]$ & 0.35 & 0.25 & 0.19 & 0.07 & $\approx$ 0 \\
\hline
$M[\chi_b(2P)]$ & 10.25 & 10.30 & - & - & - \\
\hline
$E_B[\chi_b(2P)]$ & 0.05 & $\approx$ 0 & - & - & - \\
\hline
\hline
\end{tabular}
\caption{Summary of masses and binding energies (in [GeV]) for $P-$wave 
quarkonia in the QGP as extracted from the finite-temperature 
$T-$matrix determinant, Eq.~(\ref{boundstate}).}
\label{Ptable}
\end{table}
\end{center}

%%%%%%%%%%%%%%%%%%%%%%%%%%%%%%%%%%%%%
\subsubsection{Continuum Scattering}
\label{sssec_cont}
\begin{center}
\begin{figure}[!tb]
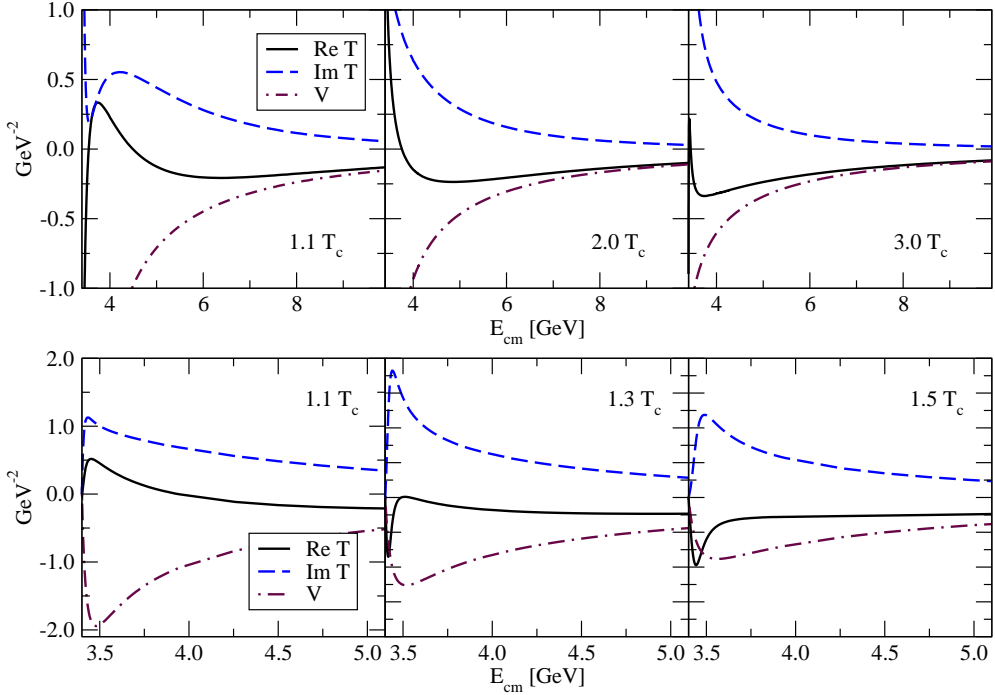

\includegraphics[width=0.8\textwidth]{TccS-contcloseup.eps}\\
\includegraphics[width=0.8\textwidth]{TccP-contcloseup.eps}
\caption{$c$-$\bar{c}$ scattering amplitude in $S-$ (up) and $P-$wave (below) above
threshold. The Born approximation to the amplitude is also shown
(dashed-dotted).}
\label{fig:Tcc-closeup}
\end{figure}
\end{center}

\begin{center}
\begin{figure}[!tb]
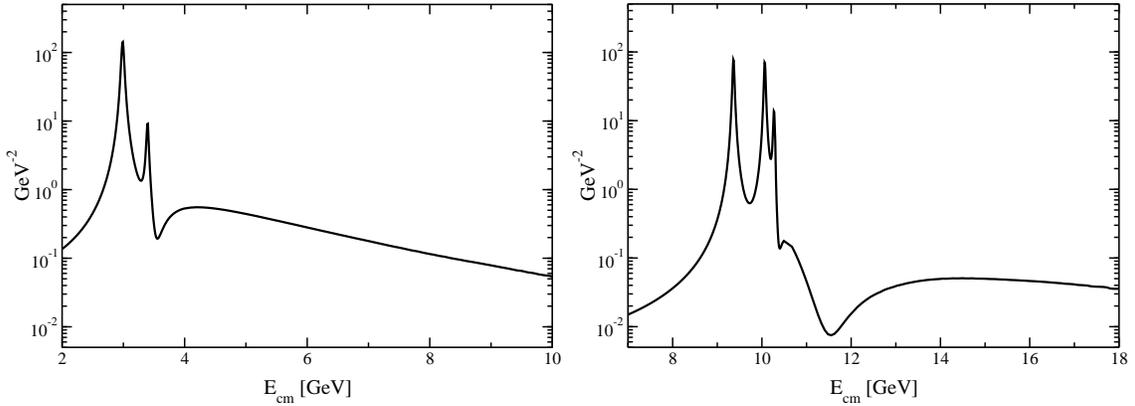

\includegraphics[width=0.45\textwidth]{ImTccS-log.eps}
\includegraphics[width=0.45\textwidth]{ImTbbS-log.eps}
\caption{Imaginary part of the $S-$wave scattering amplitude including bound and
scattering parts of the spectrum (left, charmonium; right, bottomonium) at
$T=1.1\,T_c$.}
\label{fig:ImT-log}
\end{figure}
\end{center}

%%%%%%%%%%%%%%%%%%%%%%%%%%%%%%%%%%%%%
The $T-$matrix approach also encompasses the continuum part of the 
spectrum. This is not easily appreciated in Figs.~\ref{fig:TccS} 
and \ref{fig:TbbS} because of the different scales of the narrow bound 
state signal and the amplitude above threshold. Note that the 
determinant of the transition matrix, $\det {\cal F}$, vanishes for some 
energies above threshold (cf.~Figs.~\ref{fig:TccS}-\ref{fig:detF-P}), 
possibly indicating resonant scattering in the
continuum. Indeed, as can be seen in Fig.~\ref{fig:Tcc-closeup}, the 
$T-$matrix for $c$-$\bar{c}$ $S-$ and $P-$wave scattering exhibits 
substantial correlations above threshold. The imaginary part shows a 
distorted resonant shape, which peaks at approximately the same energy 
where the real part vanishes. The non-perturbative effects of the 
$Q$-$\bar{Q}$ rescattering above threshold are evident as we compare the 
$T-$matrix to its Born approximation, $V$ (dash-dotted line in 
Fig.~\ref{fig:Tcc-closeup}). We confirm that the $T-$matrix shows
the expected behavior at high energies, i.e., the real part converges 
to the Born approximation and the imaginary part tends to vanish. 
Finally, Fig.~\ref{fig:ImT-log} displays the imaginary part of the 
$S-$wave scattering amplitude on a logarithmic scale over a wide energy 
range below and above the $Q$-$\bar{Q}$ threshold.

%%%%%%%%%%%%%%%%%%%%%%%%%%%%%%%%%%%%%%%%%%%%%%%
\subsubsection{Sensitivity to lQCD Potential}
\label{sssec_pot}
\begin{center}
\begin{figure}[!th]
\includegraphics[width=0.8\textwidth]{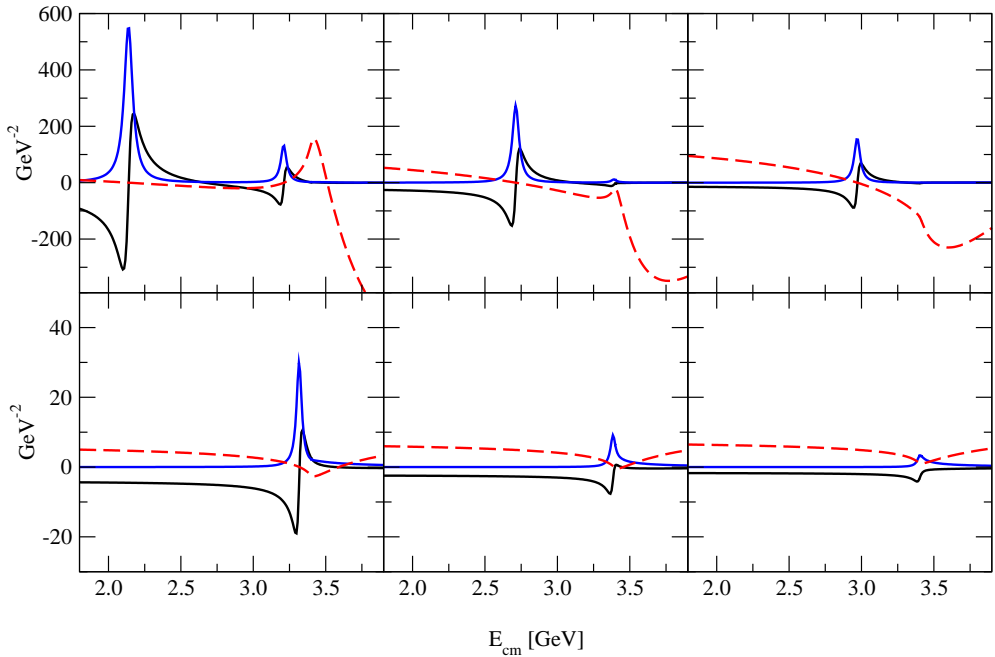}
\caption{Same as in Fig.~\ref{fig:TccS} for the potential derived from the 
lQCD internal energy of Ref.~\cite{Kaczmarek:2005gi}. From left to 
right and up to down the temperatures are
$(1.10,1.15,1.20,1.50,2.00,2.50)\,T_c$.}
\label{fig:TccS-KZ}
\end{figure}
\end{center}

%%%%%%%%%%%%%%%%%%%%%%%%%%%%%%%%%%%%%%%%%%%%%%%
In Fig.~\ref{fig:TccS-KZ} we show the $c$-$\bar c$ $S-$wave scattering 
amplitude based on the potential fitted directly to the internal energy 
data of Ref.~\cite{Kaczmarek:2005gi}. We have kept $m_c=1.7$~GeV for 
comparison with our previous results. The $T-$matrix exhibits two bound 
states with a stronger binding, as to be expected from the potential
comparison in Fig.~\ref{fig:V1}. With the same bare quark mass, the 
ground state is located at a much lower energy, $E\approx 2.2$~GeV, 
translating into a binding energy of about 1.2~GeV, in agreement with 
the results of Refs.~\cite{Alberico:2005xw,Wong:2006dz} within a
Schr\"odinger equation using the same potential. As the temperature
increases, the bound states rapidly shift to higher energies, reflecting
the rapid reduction of the potential strength at low temperatures. This trend
slows down beyond $2\,T_c$, and the ground state eventually dissolves at 
$T\simeq2.5\,T_c$. The large binding within this potential 
requires appreciable bare quark masses ($m_c\sim 2$~GeV) in order to reproduce 
the nominal position of the charmonium ground state in the vacuum. 
The strong attraction is presumably related 
to the entropy contribution to the $Q$-$\bar Q$ free energy at large
distances (cf.~Figs.~3 and 4 in Ref.~\cite{Kaczmarek:2005gi}), which 
peaks at $T_c$ and decreases steeply with temperature. The large-distance
limit of the internal energy, $U_1^{\infty}$, which inherits this 
behavior, is subtracted to generate the $Q$-$\bar Q$ potential, 
cf.~Eq.~(\ref{UminusUinf}). 
As mentioned in Sec.~\ref{ssec:Lat}, $U_1^{\infty}$ might be  
interpreted as a contribution to the in-medium quark mass,
$m_c^*(T)=m_c+U_1^{\infty}(T)/2$, i.e., a quark selfenergy contribution.
However, if no further $r$-dependence (or momentum dependence) is
considered, the simple subtraction of $U_1^{\infty}$ from the internal 
energy distorts the normalization of the potential at short distances,
where it should be described by perturbative QCD (one-gluon exchange).
A more detailed investigation of these interplays will be carried out
in future work. 

%%%%%%%%%%%%%%%%%%%%%%%%%%%%%%%%%%%%%%%%%%%%%%%%%%%%%%%%%%%%%%%%%%%%%
\section{Quarkonium Spectral Functions and Euclidean-Time Correlators}
\label{sec:Corr}
%%%%%%%%%%%%%%%%%%%%%%%%%%%%%%%%%%%%%%%%%%%%%%%%%%%%%%%%%%%%%%%%%%%%

%%%%%%%%%%%%%%%%%%%%%%%%%%%%%%%%%%%%%%%%%%%%%%%%%%%%%%%%%
\subsection{Spectral Function}
\label{ssec:Spec}
%%%%%%%%%%%%%%%%%%%%%%%%%%%%%%%%%%%%%%%%%%%%%%%%%%%%%%%%%
The $T-$matrix formalism used above can be directly applied to evaluate
mesonic spectral functions for the different quarkonium channels. The 
spectral functions encode the information on both the bound 
and scattering states in the continuum ($E>E_{th}$), similar to the
$T-$matrix. Moreover, they allow for a quantitative connection between 
the present approach and Euclidean-time correlation functions, which 
have been calculated in lattice QCD with rather high 
precision~\cite{Datta:2003ww,Petrov:2005ej}. Such a comparison has 
recently been conducted
in Ref.~\cite{Mocsy:2005qw} where the heavy-quark interaction in the
QGP has been studied by solving the bound-state problem using
a Schr\"odinger equation with either a screened Cornell-type potential
or lQCD-based internal energies (similar to the present work).
The quarkonia spectral functions were then composed of $\delta$-function
like bound states with weights determined by the decay
constant of the state and a continuum assuming free quark propagation 
with a threshold behavior taken from perturbative 
QCD~\cite{Karsch:2000gi,Alberico:2004we},
\be
\label{sigmaMP}
\sigma(\omega,T) = \sum_i 2 M_i F_i^2 \delta(\omega^2-M_i^2) + 
\frac{3}{8\pi^2} \,\omega^2\,f(\omega,E_{th}) \Theta(\omega-E_{th}) \ .
\ee
The decay constants are related to the (derivative of the) radial wave 
function at the origin for $S-$ ($P-$)wave states~\cite{Bodwin:1994jh},
while the functional form of the continuum threshold, 
given by $f(\omega,E_{th})$, depends on the specific channel 
(pseudo-/scalar, axial-/vector)~\cite{Karsch:2000gi,Alberico:2004we}. 
The threshold energies were set to $E_{th}^{c\bar{c}}=4.5$~GeV and 
$E_{th}^{b\bar{b}}=11$~GeV, based on the
phenomenological observation that no narrow mesonic resonances 
appear in the spectrum beyond this energies. 
The resulting correlation functions qualitatively reproduced the 
features observed in lQCD for the scalar channel ($\chi_{c,b}$),
whereas sizable discrepancies were found for the pseudoscalar  
and vector channels ($\eta_{c,b}$ and $J/\psi$, $\Upsilon$).

\begin{center}
\begin{figure}[!th]
\includegraphics[width=0.8\textwidth]{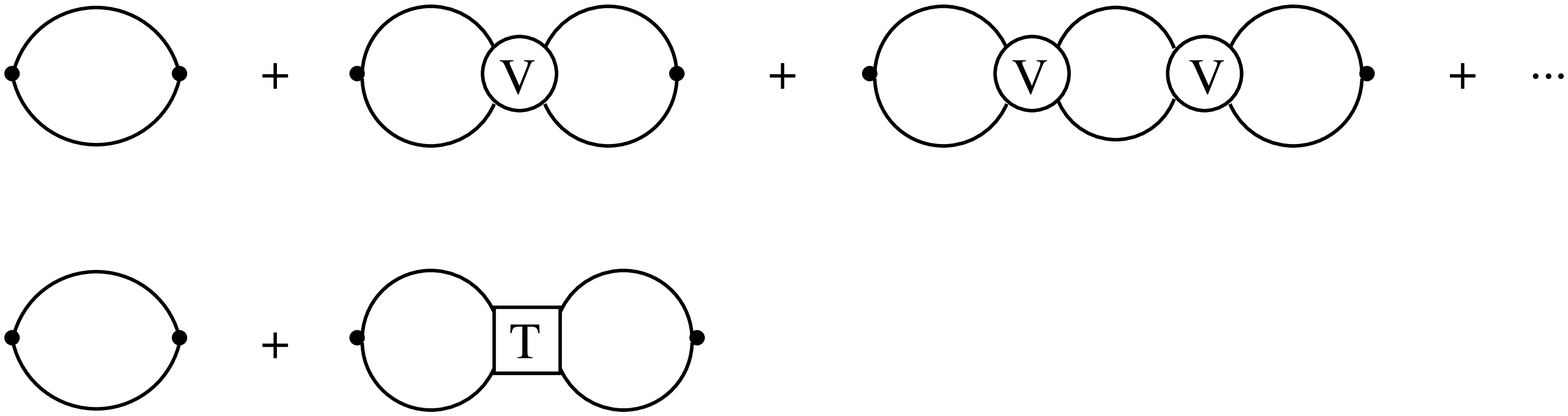}
\caption{Diagrammatic representation of the $Q$-$\bar{Q}$ correlation function. The
solid dots represent $\Gamma_M$ operators specifying different mesonic channels.}
\label{fig:spectralfun}
\end{figure}
\end{center}

In the present approach, the $Q$-$\bar{Q}$ system is interacting also 
above the $Q$-$\bar{Q}$ threshold with the same potential that generates 
the bound-state solutions, which, in particular,
accounts for the transition between the discrete and the continuum 
part of the spectrum. Nonperturbative effects
play an especially important role when reduced binding energies
drive states toward and across the two-particle threshold.
The mesonic spectral function is given by the imaginary part of the
heavy-quark two-point (current-current) correlation function in 
momentum space, $G(E,\vec{P})$, as pictorially depicted by
its perturbation series in Fig.~\ref{fig:spectralfun}. The correlation function
can be calculated from the $T-$matrix by closing the external legs with the
appropriate momentum integrations and the corresponding current operator.
Schematically, one has
\be
\label{LSschematic}
G=G^0+G^0\,T\,G^0  \ , 
\ee
where $G^0$ is the
lowest order correlation function, which represents the uncorrelated
$Q$-$\bar{Q}$ propagator in a given mesonic channel,
\bea
\label{Corr0}
G^0(E,\vec{P}= \vec{0};T) &=& i \, N_f N_c \int \frac{d^3 k}{(2\pi)^3}
\textrm{Tr} \lbrace \Gamma_M \, \Lambda_+(\vec{k}) \,
\Gamma_M \, \Lambda_-(-\vec{k}) \rbrace \, G_{Q\bar{Q}}(E;k) \, \lbrack
1-2\,f^Q(\omega_k) \rbrack \ ,
\eea
with 
$\Lambda_{\pm}(\vec{k}) = 
(\omega_k \gamma^0 - \vec{k}\vec{\gamma} \pm m_Q)/2\,m_Q$
the positive/negative energy projectors,
$\Gamma_M=(1,\gamma_5,\gamma^{\mu},\gamma^{\mu}\gamma_5)$ and $N_f\,(N_c)$
the number of flavors (colors). We take $N_f=1,N_c=3$ as in
\cite{Mocsy:2005qw} to ensure the same normalization of the lowest order
correlation function.
Eq.~(\ref{Corr0}) denotes the finite-temperature result, and we have
used the explicit decomposition of the single 
particle propagator, $S_Q$, in terms of energy projectors, which recovers 
the BbS 3D-reduction scheme used in the calculation of the $T-$matrix.

The $Q$-$\bar{Q}$ rescattering is encoded in the second (two-loop) term of
Eq.~(\ref{LSschematic}).
A proper connection has to be made between a potential description of the
interaction and the relativistic invariant amplitude entering $\Delta G$ (see
the Appendix for details). In particular, we have thus far suppressed 
the tensor structure of $T$ (and $V$) in
Dirac space. At high energies the $Q$-$\bar{Q}$ interaction should correspond 
to perturbative one-gluon exchange, which has a vector structure,
whereas at low energies lQCD finds the potential to be compatible with a 
scalar structure~\cite{Gara:1988pq}. In absence of further information, 
particularly for the
intermediate energy regime, we consider both tensor structures
alternatively and write
for the matrix elements $T_D = \bar{u}\, \tilde{\Gamma} \,u \, T \, \bar{v}
\, \tilde{\Gamma}\, v$, with $\tilde{\Gamma}=1,\gamma^{\nu}$ and $u \, (v)$ the
positive (negative) energy Dirac spinors. This leads to the following 
traces to be evaluated in Eq.~(\ref{LSschematic}):
\be
\label{tracesDirac}
\textrm{Tr}(\Gamma_M,\tilde{\Gamma}) =
\textrm{Tr} \lbrace \Lambda_+(\vec{k})\, \Gamma_M \,\Lambda_-(-\vec{k})\, 
\tilde{\Gamma} \,\Lambda_-(-\vec{k}\,') \,\Gamma_M \, \Lambda_+(\vec{k}\,') 
\, \tilde{\Gamma}
\rbrace  \ .
\ee
It turns out that they can be written in a partial wave expansion as 
performed for the $T$-matrix,
\be
\label{Tr-PWA}
\textrm{Tr}(\Gamma_M,\tilde{\Gamma}) = 
a_0^{\Gamma_M,\tilde{\Gamma}}(k,k') \, P_0(\cos \theta_{kk'})
+ a_1^{\Gamma_M,\tilde{\Gamma}}(k,k') \, P_1(\cos \theta_{kk'})
+ a_2^{\Gamma_M,\tilde{\Gamma}}(k,k') \, P_2(\cos \theta_{kk'}) \ ,
\ee
so that all angular integrations can be done analytically by using the
orthogonality of the Legendre polynomials. We thus have
\bea
\label{delta-G}
\Delta G(E;T) &=& N_f N_c \, \frac{1}{8\pi^4}
\int dk \, k^2 G_{Q\bar{Q}}(E;k) [1-2f^Q(\omega_k)]
\nonumber \\
&\times&
\int dk' \, k'^2 G_{Q\bar{Q}}(E;k') [1-2f^Q(\omega_{k'})]
{\cal T}(\Gamma_M,\tilde{\Gamma};E;k,k')  \ ,
\eea
with the kernel ${\cal T}$ given by
\bea
\label{projection}
{\cal T}(\Gamma_M,\tilde{\Gamma};E;k,k') \, &\equiv& \int d(\cos \theta_{kk'})
\textrm{Tr}(\Gamma_M,\tilde{\Gamma};k,k',\theta_{kk'}) \,
T(E;\vec{k},\vec{k}\,')
\nonumber \\
&=& 8\pi \, \lbrack 
a_0(k,k')\, T_0(E;k,k') +  a_1(k,k') T_1(E;k,k')
\rbrack  \ ,
\eea
and $a_l$ coefficients as tabulated in Table~\ref{Tr-table}.
We note that for a given channel, e.g.~pseudoscalar, in principle both the $S-$
and $P-$wave components of the $T-$matrix contribute to the correlation
function, whereas the usual spectroscopic (nonrelativistic) characterization 
of quarkonium states is based on orbital angular momentum quantum numbers
($LS$ scheme).  The (undesired) mixing of $S-$ and $P-$wave components in the
correlation function is related to the use of the $JM$ (helicity) basis 
of the $Q$-$\bar{Q}$ spectrum at high energies, in which a different partial wave
decomposition of the $T-$matrix follows. However, for the scalar and 
pseudoscalar channels the coefficient in Table~\ref{Tr-table} corresponding
to the ``natural'' partial wave is leading in the non-relativistic (heavy-quark)
expansion, whereas the other one, introducing an admixture of the ``unnatural''
partial wave, is of higher order, cf.~Table~\ref{Tr-exp-table}.
For simplicity, we shall work with the non-relativistic
approximation for the $a_l$ coefficients, which filters the appropriate partial
wave for the scalar and pseudoscalar channels (and for consistency with the
spin-averaged nature of the interaction potential we shall consider
pseudoscalar/vector, scalar/axialvector degeneracy for the correlation
functions as done for the $T-$matrix)\footnote{Note that at this level of
approximation the distinction between a scalar- or vector-like structure for $T$
is immaterial since both have the same heavy-quark limit for the $a_l$
coefficients (modulo global signs).}. We comment below on the accuracy of this
approximation.

\begin{center}
\begin{table}
\begin{tabular}{|c||c|c|}
%\hline
$\Gamma_M,\tilde{\Gamma}$ & $a_0(k,k')$ & $a_1(k,k')$ \\
\hline \hline
S,S & $\frac{k^2 k'^2}{m_q^4}$ & $-(\frac{\omega_k
\omega_{k'}}{m_q^2}+1) \frac{k k'}{m_q^2}$ \\
\hline
S,V & $4 \frac{k^2 k'^2}{m_q^4}$ & $2 \frac{k k'}{m_q^2}$ \\
\hline
PS,S & $1 + \frac{\omega_k \omega_{k'}}{m_q^2} + \frac{k^2+k'^2}{m_q^2} +
\frac{k^2 k'^2}{m_q^4}$ & $-\frac{\omega_k \omega_{k'}}{m_q^2}\frac{k
k'}{m_q^2}$ \\
\hline
PS,V & $-2 ( 1+ \frac{m_q^2-\omega_k \omega_{k'}}{m_q^2} ) - 4 (
\frac{k^2+k'^2}{m_q^2} + \frac{k k'}{m_q^2} )$ & 0 \\
\hline
V,S & $3 (1 + \frac{\omega_k \omega_{k'}}{m_q^2}) 
+ 2 \frac{\omega_k \omega_{k'}}{m_q^2} + \frac{4}{3} \frac{k k'}{m_q^2}$ &
$- ( 2 \frac{\omega_k \omega_{k'}}{m_q^2} +1 ) \frac{k k'}{m_q^2}$ \\
\hline
V,V & $-6 -4 \frac{k^2+k'^2}{m_q^2} -\frac{8}{3}\frac{k^2 k'^2}{m_q^4}$  & 
$-4 (1+ \frac{\omega_k \omega_{k'}}{m_q^2}) \frac{k k'}{m_q^2} $\\
\hline
AV,S & $-1 - \frac{\omega_k \omega_{k'}}{m_q^2} - \frac{4}{3} 
\frac{k^2 k'^2}{m_q^4}$ & $(2 \frac{\omega_k \omega_{k'}}{m_q^2} + 3) 
\frac{k k'}{m_q^2}$ \\
\hline
AV,V & $2 ( 2 \frac{\omega_k \omega_{k'}}{m_q^2} -1 
-\frac{8}{3} \frac{k^2 k'^2}{m_q^4} )$ & $- 4 \frac{\omega_k \omega_{k'}}{m_q^2} 
\frac{\omega_k \omega_{k'}}{m_q^2}$ \\
\hline
\end{tabular}
\caption{$a_l$ coefficients in a partial wave basis up to $L=1$.}
\label{Tr-table}
\end{table}
\end{center}

\begin{center}
\begin{table}
\begin{tabular}{|c||c|c|}
%\hline
$\Gamma_M,\tilde{\Gamma}$ & $a_0(k,k')$ & $a_1(k,k')$ \\
\hline \hline
S,S & ${\cal O}(k^2/m_q^2)$ & $-2 \frac{k k'}{m_q^2} + {\cal O}(k^2/m_q^2)$ \\
\hline
S,V & ${\cal O}(k^2/m_q^2)$ & $2 \frac{k k'}{m_q^2}$ \\
\hline
PS,S & $2 + {\cal O}(k^2/m_q^2)$ & ${\cal O}(k/m_q)$ \\
\hline
PS,V & $-2 + {\cal O}(k^2/m_q^2)$ & 0 \\
\hline
\end{tabular}
\caption{Lowest order of $a_l$ coefficients in a $(1/m)$ expansion.}
\label{Tr-exp-table}
\end{table}
\end{center}

%%%%%%%%%%%%%%%%%%%%%%%%%%%%%%%%%%%%%%%%%%%%%%%%%%%%%%%%%
\subsection{Euclidean-Time Correlation Functions}
\label{ssec:Eucl}
%%%%%%%%%%%%%%%%%%%%%%%%%%%%%%%%%%%%%%%%%%%%%%%%%%%%%%%%%
The Euclidean-time correlation function is defined as the thermal mesonic 
two-point correlation function in a mixed Euclidean-time-momentum 
representation~\cite{Shuryak:1993kg,Karsch:2000gi,Alberico:2004we}. 
It can be expressed in a spectral representation as an
integral transformation of the mesonic spectral function (here
$\vec{P}=\vec{0}$), 
\be
\label{EuCorr}
G(\tau,T) = \int_0^{\infty} d\omega \, \sigma(\omega,T) \, 
{\cal K} (\tau,\omega, T) \,\,\, ,
\ee
where the kernel of the transformation,
\be
\label{Kernel}
{\cal K} (\tau,\omega, T) = \frac{\cosh[\omega (\tau - \beta/2)]}{\sinh(\omega
\beta/2)} \,\,\, ,
\ee
is symmetric with respect to $\tau = \beta /2$ (and $\tau \in [0,\beta]$). The
Euclidean-time correlation function scans the full spectrum of the system.
In particular, for $\tau \to 0$ the kernel decreases rather slowly
with energy and thus the correlation function is 
dominated by contributions of the $Q$-$\bar{Q}$ continuum. 
On the other hand, for $\tau \to \beta/2$, the kernel exhibits the
maximal decrease, so that the correlation
function becomes mostly sensitive to the contribution from the 
low-energy region of the spectrum, in particular the bound states. 
To isolate the medium effects on the mesonic spectral function
from the temperature dependence introduced by the kernel it has
been proposed~\cite{Datta:2003ww,Petrov:2005ej} to normalize the correlation 
function at a given temperature to a so-called ``reconstructed" 
correlation function, 
\be
\label{recon}
G_r(\tau,T) = \int_0^{\infty} d\omega \, \sigma(\omega,T=0) \, 
{\cal K} (\tau,\omega, T) \,\,\, ,
\ee
which is obtained by replacing $\sigma (\omega,T)$ by a reference spectral
function (for instance the vacuum spectral function) and transformed
with the same finite-temperature integral kernel.  As in 
Ref.~\cite{Mocsy:2005qw}, we will first assume $\sigma(\omega,T=0)$ of the
form in Eq.~(\ref{sigmaMP}) with the vacuum input for the bound-state
part and a shape function $f(\omega,E_{th})$
given by the perturbative QCD continuum.
For the open-charm (bottom) threshold, we consider 
$E_{th}^{c\bar{c} \, (b\bar{b})}=2\,M_{D(B)}=3.74 \, (10.56)$~GeV,
but also check the sensitivity to changes in the pertinent 
free open heavy-flavor meson thresholds by using 
$E_{th}^{c\bar{c} \, (b\bar b)}=4.5\,(11)$~GeV as in Ref.~\cite{Mocsy:2005qw}.
As we shall see, the use of a simplified spectral function as in
Eq.~(\ref{sigmaMP}) may introduce spurious features in the normalized
correlation function which could mask the actual effect of the medium-modified
$Q$-$\bar{Q}$ interaction. In order to enable a more direct comparison to
lQCD evaluations, we will also normalize our results to actual spectral
functions calculated in our approach.

It is clear from Eq.~(\ref{EuCorr}) that the full energy regime of the 
spectral function figures into the calculation of the Euclidean-time correlation
function. The approximations introduced in Sec.~\ref{ssec:Spec} are expected to
be reliable up to energies above the $Q$-$\bar{Q}$ threshold, where  
non-perturbative effects from the $Q$-$\bar{Q}$ interaction prevail, as 
is already manifest from non-trivial structure in the
$T-$matrix. For higher energies in the continuum region we do not expect these
approximations to hold. However, the high energy part of the continuum should be
only relevant for $\tau \to 0$,
where the normalized correlation function approaches 1 and is not sensitive to
the evolution of the quarkonia states with the temperature.

%%%%%%%%%%%%%%%%%%%%%%%%%%%%%%%%%%%%%%%%%%%%
\subsection{Numerical Results}
\label{ssec:Cres}
%%%%%%%%%%%%%%%%%%%%%%%%%%%%%%%%%%%%%%%%%%%

%%%%%%%%%%%%%%%%%%%%%%%%%%%%%%%%%%%%%%%%%%%%%%%%%%%%%%%%%%%%
\subsubsection{Constant Heavy-Quark Mass and Small Width}
\begin{center}
\begin{figure}[!tb]
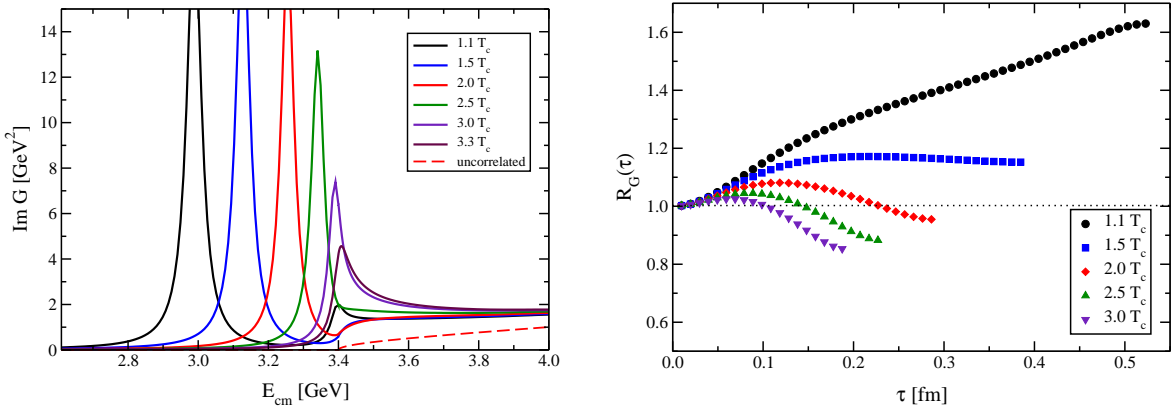

\includegraphics[width=0.45\textwidth]{ImGccS.eps}
\hspace{0.5cm}
\includegraphics[width=0.45\textwidth]{RccS-2.eps}
\caption{Left panel: Imaginary part of the correlated two-particle propagator 
(Green's function) for $c$-$\bar{c}$ $S-$wave scattering at several temperatures
and constant quark mass, $m_c$=1.7\,GeV. 
The imaginary part of the uncorrelated propagator (dashed line) is also 
shown for reference.
Right panel: Corresponding normalized mesonic correlation functions.}
\label{fig:SpecCorr-ccS}
\end{figure}
\end{center}

%%%%%%%%%%%%%%%%%%%%%%%%%%%%%%%%%%%%%%%%%%%%%%%%%%%%%%%%%%%%
%%%%%%% Charmonium S and P wave
Following our studies of the $T-$matrices, we will first investigate
mesonic spectral functions and normalized correlation
functions for a constant quark mass, and therefore the
continuum threshold is not dependent on the temperature. 
The $S-$wave charmonium spectral function is 
shown in the left panel of Fig.~\ref{fig:SpecCorr-ccS} 
for several temperatures, together with the uncorrelated (perturbative)
two-particle continuum, Eq.~(\ref{Corr0}).
As expected, the spectral function exhibits the same charmonium bound states
as found in the $T-$matrix, as well as their evolution to higher energies as
the temperature is increased. At all temperatures, the (non-perturbative) 
rescattering of the $Q$-$\bar Q$ system dynamically generates a substantial 
enhancement of strength above the $c$-$\bar{c}$ threshold relative to
the uncorrelated two-particle continuum (at $T=1.1 \,T_c$ the remnant of the
first excited state ($\psi(2S)$) is still visible). This important effect
reflects the $Q$-$\bar Q$ correlations already found for the $T-$matrix above
threshold (especially when a bound state passes into the continuum).

The corresponding normalized Euclidean correlation function for the 
same set of temperatures is displayed in the right panel of 
Fig.~\ref{fig:SpecCorr-ccS}. The normalized correlator converges to 
unity at $\tau\to 0$, which reconfirms the correct normalization
of the continuum part of the spectrum (it is also symmetric with respect 
to $\beta/2$).
At low $\tau$, where the integral in Eq.~(\ref{EuCorr}) is mostly 
dominated by the continuum region, the normalized correlator moderately 
increases, reaches a maximum, and then rapidly drops for 
$\tau$ approaching $\beta/2$, indicating a loss of strength of the 
finite temperature correlator relative to the zero-temperature one 
in the low-energy part of the spectrum. 
The temperature evolution of the correlation function is a combined result of a
decrease in binding energy of the bound states and the contribution of the
non-perturbative continuum. The sizable drop at large $\tau$ is in qualitative
agreement with the lQCD charmonium $S-$wave correlator \cite{Datta:2003ww}. The
latter, though, exhibits a smaller reduction and a weaker $\tau-$dependence,
showing appreciable deviations from unity only beyond $T=1.5\,T_c$.

\begin{center}
\begin{figure}[!th]
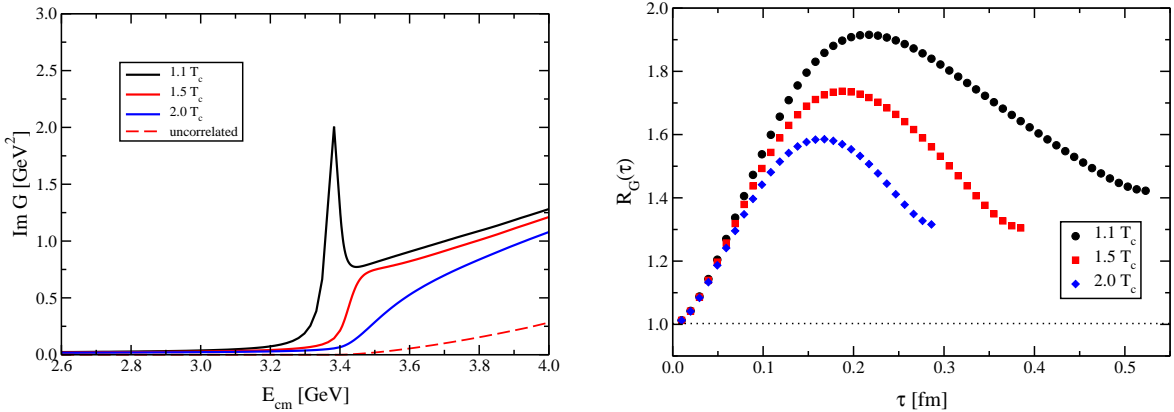

\includegraphics[width=0.45\textwidth]{ImGccP.eps}
\hspace{0.5cm}
\includegraphics[width=0.45\textwidth]{RccP-2.eps}
\caption{Same as in Fig. \ref{fig:SpecCorr-ccS} for $c$-$\bar{c}$ $P-$wave
scattering.}
\label{fig:SpecCorr-ccP}
\end{figure}
\end{center}

The charmonium spectral function and normalized correlator for $P-$wave
scattering is displayed in Fig.~\ref{fig:SpecCorr-ccP}. As was already 
discussed in
Sec.~\ref{ssec:Tmat1}, we only find a single bound state ($\chi_c$) 
just below the threshold, which rapidly melts into the continuum as 
the temperature increases. Consequently, a sizable threshold enhancement 
effect is observed. Despite the fast melting of the $P-$wave state, the
correlation function steeply rises in the 
low-$\tau$ regime, due to (i) the contribution from the non-perturbative
rescattering above threshold, and (ii) a larger threshold energy 
in the schematic vacuum spectral function entering the ``reconstructed''
correlator. 
For $\tau\to\beta/2$, the correlator levels off well above unity, 
due to the absence of an energy gap between the $P-$wave state and the 
continuum, which renders the (enhanced) continuum contribution to the 
correlator dominant even for $\tau\to\beta/2$.
The main features of our results at a given temperature are 
qualitatively in line with the lQCD $P-$wave correlators~\cite{Datta:2003ww};
however, the temperature dependence is not: our correlators attenuate
with temperature whereas the lQCD correlator increases. This appears to
be a rather direct indication that the in-medium $c$-$\bar c$ threshold
is lowered with increasing temperature.

While our results are qualitatively similar to those of 
Ref.~\cite{Mocsy:2005qw}, the following observations are in order. 
In Ref.~\cite{Mocsy:2005qw}, the increase of the correlator at low 
and intermediate $\tau$ is induced by the temperature-dependent
decrease of the continuum threshold.
Thus far, we have not
considered the temperature effect on the $Q$-$\bar Q$ threshold.
However, the non-perturbative enhancement in the spectral function
around threshold and above, which is not included in 
Ref.~\cite{Mocsy:2005qw}, turns out to be essential for a quantitative
assessment of the quarkonium correlation (and spectral) functions,
especially when the energy gap between the discrete and continuum parts 
of the spectrum is small or absent~\cite{Rapp:2006ii,Cabrera:2006nt,Wong:2006bx}.
Nevertheless, our results leave room for a lowering of the $Q$-$\bar Q$
threshold energy, since a downward shift of strength in the spectral 
function would improve (i) on the large-$\tau-$ decrease in the $S-$wave 
correlator, and (ii) on the temperature
dependence in the $P-$wave correlator, as discussed below.

\begin{center}
\begin{figure}[!th]
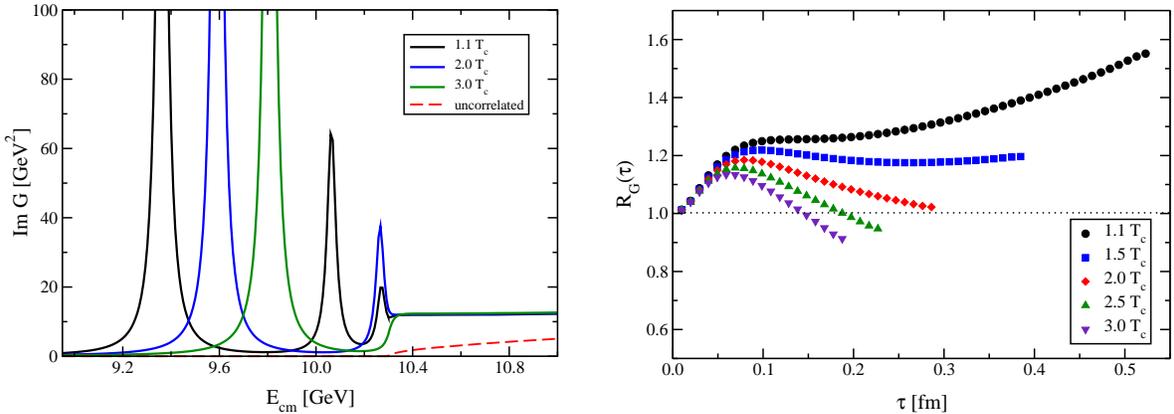

\includegraphics[width=0.45\textwidth]{ImGbbS.eps}
\hspace{0.5cm}
\includegraphics[width=0.45\textwidth]{RbbS-2.eps}
\caption{Same as in Fig. \ref{fig:SpecCorr-ccS} for $b$-$\bar{b}$ $S-$wave
scattering.}
\label{fig:SpecCorr-bbS}
\end{figure}
\end{center}

%%%%%%%%%%%% Bottomium S- and P-waves
The bottomonium spectral and correlation functions follow a similar pattern as
for the charmonium system.
The results for $S-$wave scattering are displayed in
Fig.~\ref{fig:SpecCorr-bbS}. 
At large Euclidean time, 
the correlator moderately
decreases with temperature as the two excited
bottomonia dissappear into the continuum. The contribution of the 
$\Upsilon (1S)$ state, which survives up to rather high temperatures, 
makes the correlator fall more slowly than in the charmonium case.
We find a similar enhancement at low $\tau$, partly due to the
continuum contribution with respect to the zero  
temperature case, which becomes more relevant as the bound-state
binding energies decrease with temperature.

\begin{center}
\begin{figure}[!th]
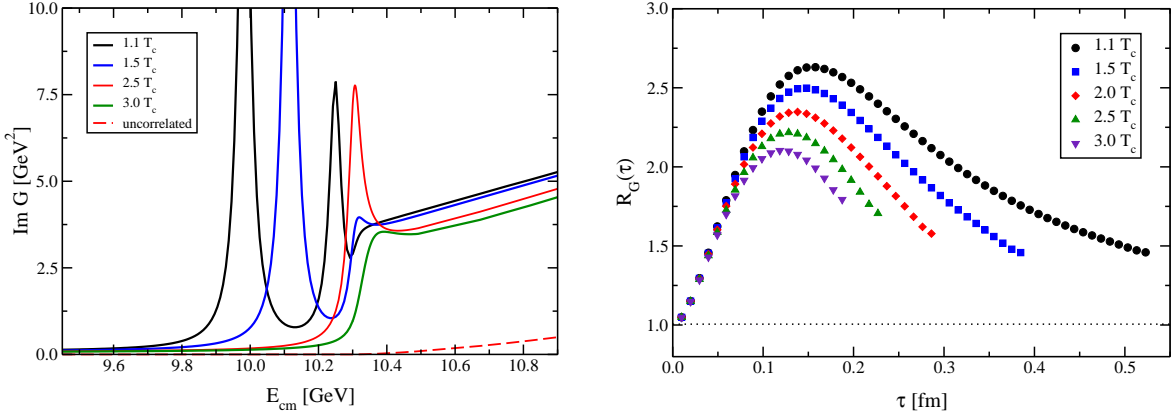

\includegraphics[width=0.45\textwidth]{ImGbbP.eps}
\hspace{0.5cm}
\includegraphics[width=0.45\textwidth]{RbbP-2.eps}
\caption{Same as in Fig. \ref{fig:SpecCorr-ccS} for $b$-$\bar{b}$ $P-$wave
scattering.}
\label{fig:SpecCorr-bbP}
\end{figure}
\end{center}

The $P-$wave bottomonium correlator, Fig.~\ref{fig:SpecCorr-bbP}, shows a
large enhancement at all $\tau$, even larger than that of the
$P-$wave charmonium correlator. A similar enhancement is 
observed in the scalar bottomonium correlator from 
lQCD~\cite{Petrov:2005ej}. As the temperature increases, the two 
bound states gradually move to higher energies and the
correlator is notably attenuated. Again, as the ground state approaches 
the continuum the nonperturbative threshold strength in the spectral 
function is the decisive source of the remaining correlator enhancement.
This reiterates the point that a comprehensive description of the 
Euclidean-time correlators should account for both the bound state 
evolution and $Q$-$\bar Q$ correlations above threshold.

%%%%%%%%%%%%%%%%%%%%%%%%%%%%%%%%%%%%%%%%%%%%%%%%%%%%%%%%%%%
\subsubsection{Sensitivity to the Reconstructed Correlator}
%%%%%%%%%%%%%%%%%%%%%%%%%%%%%%%%%%%%%%%%%%%%%%%%%%%%%%%%%%%
%%%%%%%%% Sensitivity on the threshold position for the T=0 correlator
Although our results thus far roughly reproduce some of the 
trends of the normalized 
correlators from lQCD, the $S-$wave correlator exhibits a marked $\tau$ 
dependence which is inconsistent with lQCD results. In particular,
in the low- and intermediate-$\tau$ regime, our $S-$wave correlators 
significantly increase even for temperatures just above $T_c$,
while in lQCD they are essentially unmodified until about $1.5\,T_c$. As
discussed above, one reason for this rise is the threshold mismatch 
between the finite-temperature spectral function and the vacuum one, 
which is modeled according to Eq.~(\ref{sigmaMP}).
We therefore investigate two alternative assumptions for the reconstructed 
correlator (which is used for normalization).

\begin{center}
\begin{figure}[!tb]
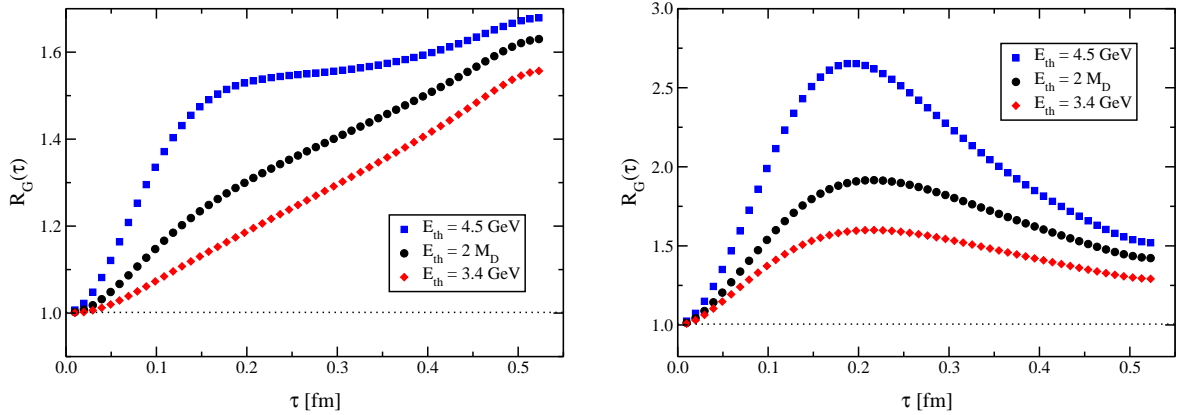

\includegraphics[width=0.45\textwidth]{RccSthres.eps}
\hspace{0.5cm}
\includegraphics[width=0.45\textwidth]{RccPthres.eps}
\caption{Normalized correlation function for $c$-$\bar{c}$ $S-$wave (left) 
and $P-$wave (right) scattering at $T=1.1 T_c$. As indicated in the 
legend, three different continuum threshold energies have been used in 
the zero temperature spectral function: $E_{th}^{c\bar{c}}=4.5$~GeV as 
in Ref.~\cite{Mocsy:2005qw}, the open-charm threshold, and
$E_{th}^{c\bar{c}}=2\,m_c$ with $m_c=1.7$~GeV.}
\label{fig:Thres-cc}
\end{figure}
\end{center}
First, we change the continuum energy threshold of the vacuum 
spectral function by varying the above values in the range
$E_{th}^{c\bar{c} \, (b\bar{b})}=3.4 - 4.5 \,(10.3-11.0)$~GeV.
In Fig.~\ref{fig:Thres-cc} 
we compare the resulting charmonium correlation function to our earlier 
result (right panels in Figs.~\ref{fig:SpecCorr-ccS} and 
\ref{fig:SpecCorr-ccP}) at the lowest temperature, $T=1.1~T_c$. 
Not surprisingly, for both $S-$ and $P-$wave scattering the low-$\tau$
rise of the amplitude is strongly reduced as the threshold energy is shifted to
lower values. 
At large $\tau$,
the $S-$wave correlator changes only slightly, as 
expected since the discrete (low-energy) part of the spectrum in the reference
spectral function is 
unchanged. However we note comparably larger changes of the $P-$wave 
correlator also at large $\tau$, since the small binding of the $\chi_c$ makes
the correlator sensitive to the continuum contribution in the entire $\tau$
domain.

\begin{center}
\begin{figure}[!th]
\includegraphics[width=0.5\textwidth]{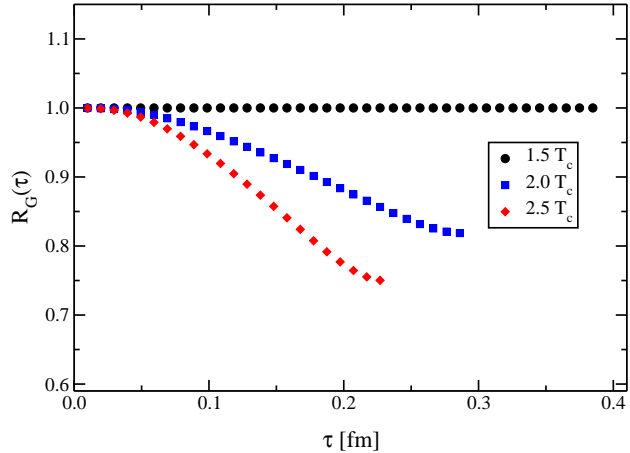}
\caption{Same as in Fig.~\ref{fig:SpecCorr-ccS} but using the
in-medium correlator at $T=1.5~T_c$ as the "reconstructed" correlator
to normalize the plotted ratio.}
\label{fig:in-med-reconst}
\end{figure}
\end{center}
Second, we make use of the observation that in the $S-$wave channel the
(normalized) lQCD correlators are essentially unchanged compared to the
reconstructed correlator up to temperatures of about $1.5\,T_c$.  
In order to remove ambiguities from the use of a simplified reference spectral
function in the reconstructed correlator, we use instead
our calculated in-medium spectral function at 1.5~$T_c$,
which incorporates the simultaneous description of bound
and scattering states. Thus, by construction, the correlator ratio at 1.5~$T_c$
is identically one, consistent with lQCD. 
The nontrivial result is now the
further temperature evolution, as displayed in  Fig.~\ref{fig:in-med-reconst}.
It turns out that the new normalized correlator shows a substantially weaker
$\tau$ dependence, decreasing with increasing temperature in a better agreement
with the lQCD correlator.

To summarize this section, we have shown that the correlator ratios
are quite sensitive to the underlying ``reconstructed" correlators
used for normalization, translating into appreciable variations  
in the absolute magnitude of the normalized correlators (especially
in the low-$\tau$ regime). 
We thus find that current discrepancies between lQCD correlation 
functions and potential-model approaches (such as in the previous 
Section, as well as in earlier works~\cite{Mocsy:2005qw}) do not
necessarily invalidate the latter, especially if one recalls the
present additional uncertainty of how to define a $Q$-$\bar Q$ potential 
from the lattice free/internal energy~\cite{Wong:2004zr,Wong:2006bx}.

%%%%%%%%%%%%%%%%%%%%%%%%%%%%%%%%%%%%%%%%%%%%%%%%%%%%%%%%%%%
\subsubsection{In-Medium Heavy-Quark Masses and Widths}
%%%%%%%%%%%%%%%%%%%%%%%%%%%%%%%%%%%%%%%%%%%%%%%%%%%%%%%%%%%

%%%%%%%%%% Spectral functions and correlators with quark effective masses

\begin{center}
\begin{figure}[!th]
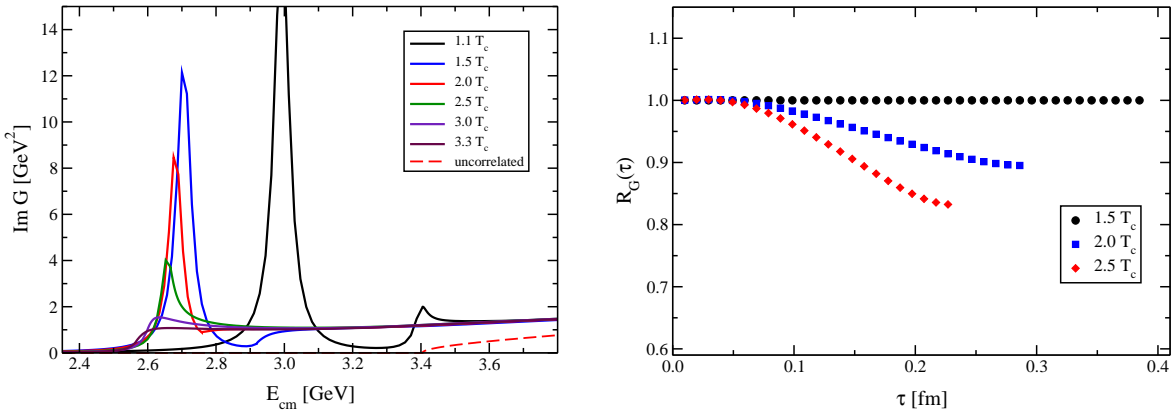

\includegraphics[width=0.45\textwidth]{ImGccSmstar.eps}
\hspace{0.5cm}
\includegraphics[width=0.45\textwidth]{RccSmstar-15Tc.eps}
\caption{Left: Imaginary part of the correlated two-particle propagator (Green's
function) for $c$-$\bar{c}$ $S-$wave scattering at several temperatures. The
temperature dependent effective quark mass, $m_q^*(T)$, has been used in the
calculation. Right: Corresponding normalized mesonic correlation function at
several temperatures.}
\label{fig:SpecCorr-ccS-mstar}
\end{figure}
\end{center}

We finally consider the effect 
of in-medium properties of the interacting heavy quarks. 
First, we incorporate an effective in-medium quark mass as extracted 
from the large-distance plateau of the internal energy, 
$m_Q^*(T) = m_Q + U_1^{\infty}(T)/2$,
and we neglect a possible momentum dependence of this correction. We note that
in our approach an effective reduction of the heavy-quark mass does not only
modify the $Q$-$\bar Q$ threshold energy ($E_{th}=2\,\omega_{q=0} = 2\,m_Q^*$), but
also figures into the two-particle propagator as a selfenergy contribution 
and therefore is iterated in the scattering equation. We again adjust the 
bare quark mass to recover the vacuum mass of the quarkonium ground state at 
$1.1\,T_c$.  The charmonium spectral function for $S-$wave scattering
(left panel of Fig.~\ref{fig:SpecCorr-ccS-mstar}) shows a sizable downward
shift of the bound state strength, which for higher temperatures
is compensated by the reduction in the
binding energy. As expected, the dissociation temperature of the ground state
is smaller, $T_{diss}\approx 2.5\,T_c$, as compared to the calculation with a
fixed quark mass. The Euclidean correlator (right panel in 
Fig.~\ref{fig:SpecCorr-ccS-mstar}), which we normalize to the 1.5~$T_c$ 
spectral function as discussed in the
previous section, 
has a rather mild 
temperature dependence, with an attenuation of the order of 10-15\%, 
in a closer agreement with the $S-$wave lQCD correlator. 
Thus, for $T\ge1.5T_c$, the $T$-dependent (decreasing) threshold 
improves the agreement with the lattice calculations. 
On the other hand, a strong $T$ dependence as dictated by the large-distance
behavior of the internal energy at $T\le1.5T_c$ would not be supported the 
lQCD correlator (a recent analysis of
charmonium properties in quenched lQCD shows a temperature dependent charmonium
mass reduction of about 6\% at $2\,T_c$ \cite{Iida:2006mv}).
For more stringent conclusions the present uncertainties in the extraction 
of the heavy-quark potential, as well as the assumption
of a 3-momentum independent mass correction (which upsets the perturbative
normalization of the potentials at short distances and therefore represents
an upper estimate of the in-medium mass effect), need to be scrutinized.

\begin{center}
\begin{figure}[!th]
\includegraphics[width=0.5\textwidth]{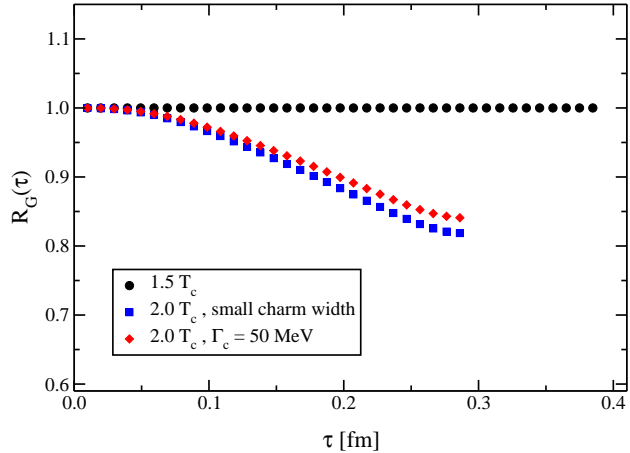}
\caption{$S-$wave charmonium correlator with a charm width of 50~MeV, as
compared to the narrow quark calculation.}
\label{fig:finitewidtheffect}
\end{figure}
\end{center}

We furthermore study the impact of finite quarkonium widths, which are
an essential ingredient in the phenomenology of charm and bottom in
heavy-ion collisions. As suggested by recent analysis of heavy-quark 
diffusion in a QGP~\cite{vanHees:2004gq,vanHees:2005wb}, as well as
parton-induced break-up reactions of charmonia~\cite{Grandchamp:2003uw},
their widths are expected to be of the order of 100~MeV at temperatures
around $1.5\,T_c$. We can easily incorporate such effects by dressing the 
charm quarks with an imaginary selfenergy in the two-particle propagator, 
Eq.~(\ref{BbS}).
The results for the $S-$wave charmonium correlation function in the fixed quark
mass approach are shown in
Fig.~\ref{fig:finitewidtheffect}, where a charm width of 50~MeV (generating 
charmonium widths of $\sim$100~MeV) has been used, in comparison with the 
narrow-width limit. At $2\,T_c$ the Euclidean correlator is modified
by only a few percent. Especially in view of other current 
uncertainties, the correlators are rather insensitive to
phenomenologically relevant magnitudes of the quarkonium decay widths;
this may after all not be surprising since $\Gamma_\psi$ is very much smaller
than the charmonium mass, and also appreciably smaller than the 
typical temperature.

%%%%%%%%%%%%%%%%%%%%%%%%%%%%%%%%%%%%%%%%%%%%%%%%%%%%
\section{Conclusions and Outlook}
\label{sec:Concl}
%%%%%%%%%%%%%%%%%%%%%%%%%%%%%%%%%%%%%%%%%%%%%%%%%%%%

In the present article we have evaluated spectral properties of
heavy quark-antiquark interactions (charm and bottom) in the
Quark-Gluon Plasma within a $T-$matrix approach which allows for
a comprehensive treatment of bound and scattering states. 
The basic interaction was taken to be 
a two-body potential which, following earlier works, has been 
identified with the heavy-quark internal energy evaluated in 
thermal lattice QCD.  
The finite-temperature $T-$matrices reconfirmed the survival of
ground-state ($S-$wave) bound states for temperatures well 
above $T_c$ as found in earlier calculations, up to $\sim 2.8\,T_c$  
($>3.5\,T_c$) for $\eta_c$/$J/\psi$ ($\eta_b$/$\Upsilon$),
as well as for $P-$wave and excited bottomonia.
The dissociation mechanism is characterized by a bound state passing 
through the $Q$-$\bar Q$ threshold, at which point a strong reduction 
and broadening of the (imaginary part of the) $T-$matrix is observed. 
Nevertheless, the $Q$-$\bar Q$ system remains strongly correlated in the 
continuum, indicated by resonant-like structures in the $T-$matrix 
(albeit at much reduced magnitude) which deviate from the Born 
approximation up to energies of 1-2\,GeV above $Q$-$\bar Q$ threshold.

We have proceeded to calculate $Q$-$\bar Q$ current-current correlation 
functions, which, in the timelike regime, follow from the 
$T-$matrix by folding with the $Q$-$\bar Q$ propagator. The imaginary
part of the correlation function (spectral function) confirmed 
the importance of threshold effects in the dissolution
mechanisms, resulting in large (nonperturbative) enhancements over
the perturbative form of the continuum. These, in turn, give
substantial contributions to the Euclidean correlators primarily at 
large and intermediate time, $\tau$, which cannot be neglected 
in  quantitative comparisons to (and interpretations of) lattice QCD 
results. Assuming constant heavy-quark masses, some qualitative
features of the lQCD correlators (large-$\tau$ decrease in the
$S-$waves, overall increase in the $P-$waves) could be reproduced.
However, the magnitude of the signal in the $S-$wave channels, as well as 
the temperature dependence in the $P-$wave channels, were inconsistent
with lQCD.  

We have found an appreciable sensitivity of the Euclidean correlator 
ratios to the so-called ``reconstructed" (reference) spectral function
used for normalization (usually taken as a vacuum form). 
The choice of shape and onset of the continuum introduces 
$\tau-$dependencies in the normalized correlator, which may render 
the identification of medium effects in the calculated spectral functions
more difficult. E.g., when normalizing to a reference spectral function
calculated at $1.5\,T_c$ for the $S-$wave $c$-$\bar c$ system
(as suggested by the corresponding lQCD result), the 
lQCD correlators for higher temperatures can be much better reproduced. 
We have also inferred that the lQCD correlators favor a temperature-dependent
decrease of the heavy-quark mass ($Q$-$\bar Q$ threshold), which pushes spectral
strength to lower energies and improves on the large-$\tau$ and
temperature dependence of the correlators.

Our approach furthermore allows for establishing a closer connection to 
quarkonium phenomenology in heavy-ion collisions by incorporating 
finite width effects. E.g., when implementing in-medium heavy-quark
widths of $\sim$50\,MeV in the two-particle propagator of the scattering
equation (inducing a charmonium width of $\sim$100~MeV as suggested 
by phenomenology), we find only few-percent changes in the Euclidean 
correlators, which are superseded by other current uncertainties.

In conclusion, our results suggest that lQCD-based potential approaches,
when consistently implementing both bound and scattering states in a
non-perturbative scheme, are a 
viable means to quantitatively interpret the rather precise lQCD
computations on Euclidean correlation functions, and thus evaluate 
the properties of quarkonium spectral functions in the QGP.
Significant uncertainties still reside in the extraction of an
appropriate $Q$-$\bar Q$ potential, as well as in the determination 
of the in-medium open-charm and -bottom masses.
Once quantitative agreement between model calculations and lQCD
correlators has been established, applications to high-energy heavy-ion 
collisions will subject the theoretical results to experimental tests.
Hopefully this will pave the way to progress on the long-standing challenge 
of connecting heavy-quarkonium observables to properties 
of the finite-temperature QCD phase transition.

\vspace{0.5cm}

\noindent
{\bf Acknowledgments} \\
We thank M.~Mannarelli and F.~Zantow for providing us with their
(fits to) lattice QCD results. We also acknowledge useful
discussions with M.~Mannarelli and H.~van Hees.
One of us (DC)  thanks Ministerio de Educaci\'on y Ciencia (Spain) for support
through a postdoctoral fellowship. One of us (RR) has been supported in part 
by a U.S.~National Science Foundation CAREER Award under
Grant No. PHY0449489.

%\bibliographystyle{apsrev}
%\bibliography{hadrons}

\appendix

\section{3-dimensional reduction of the Bethe-Salpeter equation}
The Bethe-Salpeter equation for $Q$-$\bar{Q}$ scattering with the
Blankenbecler-Sugar (BbS) three-dimensional reduction of the two-particle
propagator \cite{Blankenbecler:1965gx,Machleidt:1989tm} 
reads, in the CM frame,
\be
\label{ap:BS}
{\cal M}(E;\vec{q}\,',\vec{q}) = {\cal V}(\vec{q}\,',\vec{q})
+ \int \frac{d^3k}{(2\pi)^3} {\cal V}(\vec{q}\,',\vec{k}) \,
\frac{m^2}{\omega_k} 
\frac{\Lambda_+(\vec{k}) \, \Lambda_-(-\vec{k})}{s/4-\omega_k^2+i\epsilon}
\, {\cal M}(E;\vec{k},\vec{q}) \ ,
\ee
where the invariant amplitudes ${\cal M}$ and ${\cal V}$ are actually operators
(truncated amplitudes) which act in the direct product of the Dirac spaces of
each fermion. The BbS scheme, originally formulated for the nucleon-nucleon
($NN$) interaction,
exploits the following decomposition of the 
single-particle propagator in terms of positive- and negative-energy states,
\be
\label{SF}
S_F(k^0,\vec{k}) = \frac{m}{\omega_k}
\frac{\Lambda_+(\vec{k})}{k^0-\omega_k+i\eta}
-
\frac{m}{\omega_k}
\frac{\Lambda_-(-\vec{k})}{k^0+\omega_k-i\eta} \ . 
\ee
Consequently, the full four-dimensional two-particle propagator,
$i\,S_Q(k+P/2)\,S_{\bar{Q}}(k-P/2)$, is replaced by the following function
\be
\delta(k^0) \, \frac{m^2}{\omega_k} 
\frac{\Lambda_+(\vec{k}) \, \Lambda_-(-\vec{k})}{s/4-\omega_k^2+i\epsilon} \ ,
\ee
which has the same discontinuity across the right-hand cut and puts the quark
(anti-quark) on the positive (negative) energy shell, suppressing virtual
anti-quark (quark) contributions. Note that since both fermions are equally
off-shell the energy transfer at the interaction is zero (BbS neglects
retardation effects), and this allows for a description in terms of a static
potential, $V(\vec{q}\,',\vec{q})$, as done for instance in the Boson Exchange
Model of the $NN$ interaction \cite{Machleidt:1989tm}.

\begin{center}
\begin{figure}[!tb]
\includegraphics[width=0.7\textwidth]{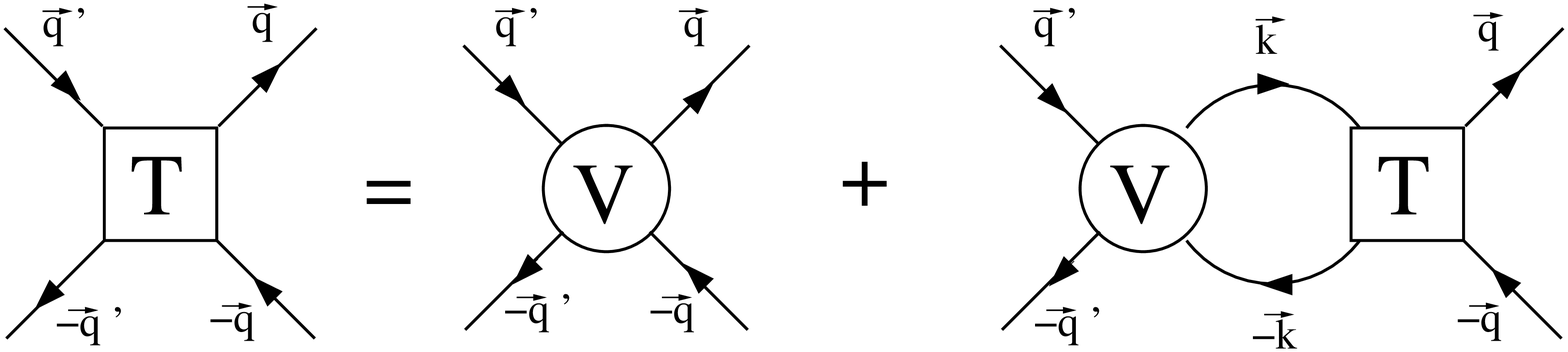}
\caption{CM kinematics of the Bethe-Salpeter equation.}
\label{fig:BSkinematics}
\end{figure}
\end{center}

One can take matrix elements in Eq.~(\ref{ap:BS}) between the
appropriate Dirac spinors, $\tilde{T}[\tilde{V}]\equiv \bar{u}(\vec{q}) \,
\bar{v}(-\vec{q}\,') {\cal M}[{\cal V}] u(\vec{q}\,') v(-\vec{q})$ (see 
Fig.~\ref{fig:BSkinematics} for kinematics), and then Eq.~(\ref{ap:BS}) can 
be rewritten as (helicity indices omitted)
\be
\label{ap:BS2}
\tilde{T}(E;\vec{q}\,',\vec{q}) = \tilde{V}(\vec{q}\,',\vec{q})
+ \int \frac{d^3k}{(2\pi)^3} \tilde{V}(\vec{q}\,',\vec{k}) \,
\frac{m^2}{\omega_k} \frac{1}{s/4-\omega_k^2+i\epsilon}
\, \tilde{T}(E;\vec{k},\vec{q}) \ , 
\ee
where we have used the following representation of the energy projectors
\bea
\label{projector2}
\Lambda_+(\vec{k}) &=& \frac{\sum_{\lambda} u_{\lambda}(\vec{k})\,
\bar{u}_{\lambda}(\vec{k})}{2\,m} \ ,
\nonumber \\
\Lambda_-(\vec{k}) &=& \frac{-\sum_{\lambda} v_{\lambda}(-\vec{k})\,
\bar{v}_{\lambda}(-\vec{k})}{2\,m} \ .
\eea
The connection between $\tilde{T},\tilde{V}$ and the actual (static) potential
$V$ and the $T-$matrix in Eq.~(\ref{LS}) can be derived by considering a
tensor structure for $\tilde{V}$ and performing a non-relativistic 
reduction of the resulting amplitude (for
instance consider $\tilde{V}$ given by the Yukawa scalar-meson exchange
amplitude, which can be fully derived from the Lagrangian, ${\cal L}_S = g_S\,
\bar{\Psi} \Psi \phi$). 
It turns out that 
$\tilde{V}(\vec{q}\,',\vec{q}) = V(\vec{q}\,',\vec{q}) + {\cal O}(q^2/m^2)$,
with $V$ related to the corresponding potential in coordinate space by
\be
\label{Fourier-ktop}
V(r) = \frac{1}{(2\pi)^3} \int d^3 k \, e^{i \vec{k} \vec{r}} \, V(\vec{k}) \ ,
\ee
and $\vec{k}=\vec{q}\,'-\vec{q}$. The partial wave decomposition of the
potential (and of $T$) is given by
\be
\label{PWA}
V(\vec{q}\,',\vec{q})=4\pi \sum_l (2l+1)\,V_l(q',q)\,P_l(\cos\theta_{q'q}) \ ,
\ee
and then Eqs.~(\ref{LS},\ref{pot-Fourier-projected}) follow.
The temperature dependence is accounted for by introducing a $[1-2f^Q]$
factor for each two-particle loop, and the quark selfenergy enters the
two-particle propagator by the replacement 
\be
\label{finitewidth}
[s/4-\omega_k^2+i\epsilon]^{-1} \longrightarrow 
[s/4-\omega_k^2-2 i\,\omega_k \textrm{Im}\,\Sigma]^{-1}
\ee
in Eq.~(\ref{ap:BS2}), 
with $\omega_k$ satisfying the dispersion relation in Eq. (\ref{disp}).

\end{document}